\documentclass[preprint,superscriptaddress,nofootinbib]{revtex4-1}
\usepackage[dvips]{graphicx}
\usepackage{amsmath,amssymb,color,colortbl}
\def\beq#1\eeq{\begin{equation}#1\end{equation}}
\def\bal#1\eal{\begin{align}#1\end{align}}
\newcommand{\di}[1]{\overline{#1}}

\newcommand{\MNS}{{\text{MNS}}}

\newcommand{\Pl}{{\text{Pl}}}

\newcommand{\eV}{{\text{eV}}}

\newcommand{\GeV}{{\text{GeV}}}

\newcommand{\BR}{\text{BR}}
\newcommand{\Br}{\text{Br}}

\newcommand{\U}{{\text{U}}}
\newcommand{\SU}{{\text{SU}}}

\begin{document}

\preprint{OU-HET 993}
\preprint{UT-HET 130}

\title{
New model for radiatively generated Dirac neutrino masses
and
lepton flavor violating decays of the Higgs boson
}

%%%%%%%%%%%%%%%%%%%%%%%%%%%%%%%%%%%%%%%%%%%%%%%%%%%%%%%%%%%%%%%%%%%%%%
\author{Kazuki Enomoto}
\email{kenomoto@het.phys.sci.osaka-u.ac.jp}
\affiliation{
Department of Physics,
Osaka University,
Toyonaka,
Osaka 560-0043, Japan
}
%%%%%%%%%%%%%%%%%%%%%%%%%%%%%%%%%%%%%%%%%%%%%%%%%%%%%%%%%%%%%%%%%%%%%%
\author{Shinya Kanemura}
\email{kanemu@het.phys.sci.osaka-u.ac.jp}
\affiliation{
Department of Physics,
Osaka University,
Toyonaka,
Osaka 560-0043, Japan
}
%%%%%%%%%%%%%%%%%%%%%%%%%%%%%%%%%%%%%%%%%%%%%%%%%%%%%%%%%%%%%%%%%%%%%%
\author{Kodai Sakurai}
\email{Address after April 2019, Institute for Theoretical Physics, Karlsruhe Institute of Technology, D-76128 \linebreak Karlsruhe, Germany; kodai.sakurai@kit.edu
}
\affiliation{
Department of Physics,
University of Toyama,
3190 Gofuku,
Toyama 930-8555, Japan
}
%%%%%%%%%%%%%%%%%%%%%%%%%%%%%%%%%%%%%%%%%%%%%%%%%%%%%%%%%%%%%%%%%%%%%%
\author{Hiroaki Sugiyama}
\email{shiro324@gmail.com}
\affiliation{
Center for Liberal Arts and Sciences,
Toyama Prefectural University,
Toyama 939-0398, Japan
}
%%%%%%%%%%%%%%%%%%%%%%%%%%%%%%%%%%%%%%%%%%%%%%%%%%%%%%%%%%%%%%%%%%%%%%%

%%%%%% date %%%%%%%
%\date{\today}
%%%%%%%%%%%%%%%%%%%

\begin{abstract}
 We propose a new mechanism to explain neutrino masses
with lepton number conservation,
in which the Dirac neutrino masses are
generated at the two-loop level
involving a dark matter candidate.
 In this model,
branching ratios of lepton flavor violating decays of the Higgs boson
can be much larger than those of lepton flavor violating decays of
charged leptons.
 If lepton flavor violating decays of the Higgs boson are observed
at future collider experiments
without detecting lepton flavor violating decays of charged leptons,
most of the models previously proposed for tiny neutrino masses are excluded
while our model can still survive.
 We show that the model can be viable
under constraints from current data for neutrino experiments,
searches for lepton flavor violating decays of charged leptons
and dark matter experiments.

\end{abstract}

\maketitle

%%%%%%%%%%%%%%%%%%%%%%%%%%
%%%  Introduction >>>  %%%
%%%%%%%%%%%%%%%%%%%%%%%%%%
\section{Introduction}

%(Overview, SM)
Although the Standard Model (SM) is consistent with the current data of collider experiments, 
there are still mysterious phenomena which cannot be explained in the SM, such as the origin of neutrino masses,  the nature of dark matter and the baryon asymmetry of the Universe. 
To explain these phenomena by extending the SM is one of the central interests of today's high energy physics. 
Various models and mechanisms have also been proposed.

%(Beyond the SM: Neutrino mass problem, DM problem, baryogenesis)

For the origin of neutrino masses, many new models have been studied along with the idea of the seesaw mechanism, which explains Majorana-type tiny neutrino masses by introducing new heavy particles, such as right-handed neutrinos~\cite{ref:seesaw,Schechter:1980gr}, an additional isospin triplet scalar field~\cite{Schechter:1980gr,ref:HTM} and isospin triplet fermions~\cite{Foot:1988aq}.  
There is also an alternative scenario where 
tiny neutrino masses are generated by quantum effects. 
The first model along this line was proposed by Zee~\cite{Zee:1980ai}, in which one-loop effects due to an additional Higgs doublet field and a charged singlet scalar field yield Majorana-type tiny neutrino masses.   
There have been many variation models~\cite{Zee:1985id,Babu:1988ki,Cheng:1980qt,ref:GNR,ref:KNT,ref:AKS,ref:Ma}, some of which introduce an unbroken discrete symmetry in order not only to forbid tree-level generation of neutrino masses but also to guarantee the stability of extra particles in the loop so that the lightest one can be identified as a dark matter candidate~\cite{ref:GNR,ref:KNT,ref:AKS,ref:Ma}.    
In Ref.~\cite{ref:AKS}, an extended scalar sector for inducing neutrino masses at the three loop level with a dark matter candidate is also used to cause the strongly first order electroweak phase transition, which is required for successful electroweak baryogenesis~\cite{Kuzmin:1985mm}.    
In addition, models which generate Dirac-type tiny neutrino masses by quantum effects have also been proposed in Refs.~\cite{ref:1loopDirac,Gu:2007ug,Kanemura:2017haa}.  
In Ref.~\cite{Kanemura:2017haa}, introducing right-handed neutrinos with an odd quantum number under a new discrete symmetry, Dirac-type tiny neutrino masses are generated at the two-loop level.
This model also has a dark matter candidate and can realize the strongly first order phase transition.

%(Higgs LFV decay and classification of neutrino models)
In Ref.~\cite{Kanemura:2015cca}, a class of models in which Majorana-type tiny neutrino masses are generated by quantum effects has been comprehensively studied by using flavor structures of induced neutrino mass matrices.  
Classification of models to generate Dirac-type neutrino masses has also been performed in Ref.~\cite{Kanemura:2016ixx}. 

%(Higgs mu tau in the LHC)
Several years ago, anomaly for a lepton flavor violating (LFV) decay process of the Higgs boson $h\to\mu\tau$ at the LHC was reported by ATLAS~\cite{Aad:2015gha} and CMS~\cite{Khachatryan:2015kon,CMS:2016qvi}, although it disappeared soon later~\cite{Sirunyan:2017xzt}. 
Motivated by this anomaly, the authors of Ref.~\cite{Aoki:2016wyl} examined in a systematic way what kind of models for neutrino masses can predict a significant amount of signals for $h\to\mu\tau$. 
It was shown that most of the proposed models radiatively generating Majorana-type neutrino masses and Dirac-type neutrino masses, as well as minimal models of Type-I, II and III seesaw mechanisms are excluded if the signal of LFV decays of the Higgs boson is observed at future collider experiments without detecting LFV process for charged leptons. 
They also found that only a few models, in which Dirac-type neutrino masses are generated radiatively, may not be excluded even in this case.

%(In this letter, ...)
In this paper, we concretely build one of such models, where additional scalar fields as well as right-handed fermions are introduced with even or odd charge under new discrete symmetries, so that Dirac-type tiny neutrino masses are generated at the two-loop level and a dark matter candidate is also contained. 
The branching ratio for LFV decays of the Higgs boson is not too small in spite of the stringent constraints from LFV processes for charged lepton decays. 
We will show that the model can be viable
under the constraints from current data for neutrino experiments,
searches for flavor violating decays of charged leptons
and dark matter experiments.

 %(Summary and organization of this letter)
This paper is organized as follows. In Sec.~\ref{sec:Model}, we define our model and introduce new fields and symmetries. 
In Sec.~\ref{sec:Neutrino Mass}, we give the formula of neutrino mass matrix which is generated at two-loop level.
In Sec.~\ref{sec:Lepton Flavor Violation}, we consider the LFV processes $\ell \to \ell^\prime \gamma, h \to \ell \ell^\prime~\text{and}~\ell_m\to\overline{\ell}_n\ell_p\ell_q$.
In Sec.~\ref{sec:Dark Matter}, we show formulae of the thermally averaged cross sections of annihilation processes of the dark matter and the relic abundance.
In Sec.~\ref{sec:Benchmark Scenarios and Numerical Evaluation}, we present two benchmark scenarios and give numerical results of various phenomena in Secs.~\ref{sec:Neutrino  Mass},~\ref{sec:Lepton Flavor Violation} and~\ref{sec:Dark Matter}. The first scenario is for the normal ordering mass hierarchy of neutrinos, and the second one is for the inverted ordering one. 
Conclusions are shown in Sec.~\ref{sec:Conclusions}. Some formulae are presented in Appendices.

%%%%%%%%%%%%%%%%%%%
%%%  Model >>>  %%%
%%%%%%%%%%%%%%%%%%%
\section{Model}
\label{sec:Model}
%(Symmetries) 

In our model,
fields listed in Table~\ref{table:particle list}
are added to the SM ones.
We impose the conservation of the lepton number $L$  to our model.
Gauge singlet right-handed fermions
$\nu_{iR}^{}$~$(i = 1, 2, 3)$
have $L = 1$, which compose three Dirac neutrinos with left-handed neutrinos
$\nu_{\ell L}^{}$~$(\ell = e, \mu, \tau)$
of the SM lepton doublet fields
$L_\ell = (\nu_{\ell L}^{},\, \ell_L^{})^T$.
On the other hand,
lepton numbers of the other gauge singlet fermions
$\psi_{aR}^{}$~$(a = 1, 2, 3)$ are zero.
They have Majorana mass terms, $\frac{1}{2}M_{\psi_a}\overline{\psi_{aR}^c} \psi_{aR}$,
without breaking the lepton number conservation,
while Majorana mass terms of $\nu_{iR}^{}$ are forbidden.
If neutrinos have Yukawa interactions
$(Y_\nu)_{\ell i} \di{L_\ell}\, \phi^c \nu_{iR}^{}$
with the SM Higgs doublet field
$\phi = (\phi^+ , \phi^0)^T$,
masses of Dirac neutrinos can be generated
with the vacuum expectation value $\langle \phi^0 \rangle$.
However,
required values of Yukawa coupling constants $(Y_\nu)_{\ell i}$
for tiny neutrino masses
seem to be unnaturally small.
Thus,
we impose a softly broken discrete symmetry $(Z_2')$ to our model
in order to forbid tree-level Yukawa interaction of neutrinos,
where $\nu_{iR}^{}$ are odd under $Z_2'$
while fields in the SM are even.
Assignments of $Z_2^\prime$ quantum number to the new fields
are shown in Table~\ref{table:particle list}.
Although neutrino masses in the lagrangian are forbidden
by $Z_2^\prime$,
they can be generated at the loop level
via the soft breaking effect in the scalar sector.
Throughout this paper,
we take the basis where $\ell$, $\nu_{iR}^{}$, and $\psi_{aR}^{}$
are mass eigenstates.

Four new scalar fields~($\Phi$, $s_1^+$, $\eta$, and $s_2^+$)
are involved in our model
in addition to the Higgs doublet field $\phi$ of the SM\@.
Both of $s_1^+$ with $L = -2$ and $s_2^+$ with $L = -1$ are
$\SU(2)_L$-singlet fields with $Y = 1$.
On the other hand,
$\Phi = (\Phi^{++}, \Phi^+)^T$ with $Y = 3/2$
and
$\eta = (\eta^+, \eta^0)^T$ with $Y = 1/2$
are $\SU(2)_L$-doublet fields.
The doublet field $\Phi$ has $L = -2$,
and the even parity under $Z_2^\prime$ is assigned to $\Phi$.\footnote{
Actually,
the $Z_2^\prime$ parity of $\Phi$ is irrelevant to our study in this article
so that the odd-parity is also acceptable for $\Phi$.}
Although $\eta$ belongs to the same representation as $\phi$
under the SM gauge symmetry,
$\eta$ has $L = -1$ in contrast with $L = 0$ for $\phi$.
 We restrict ourselves to the case
where $\eta^0$, the neutral component of $\eta$, does not have a vacuum expectation value
in order to keep the lepton number conservation.
 The other new scalar fields do not also have vacuum expectation values
because they are electrically charged.

 Apart from $Z_2'$, an accidental unbroken discrete symmetry $(Z_2)$ appears in our model
due to the lepton number conservation,
Majorana mass terms of $\psi_{aR}^{}$ and some of new Yukawa interactions,\footnote{
 These Majorana mass terms, $Y_1$ and $Y_2$ terms in Eq.~(\ref{eq:Yukawa_int}) explicitly break
$\U(1)_{L + 2J}$ into its $Z_2$ subgroup.
}
where the parity is given by $(-1)^{L + 2J}$.
 Three fields~($\psi_{aR}^{}$, $\eta$, and $s_2^+$)
are odd under $Z _2$.
 The lightest $Z_2$-odd particle is stable.
 If $\psi_{aR}^{}$ or $\eta^0$ is the lightest one,
it can be a dark matter candidate.

\begin{table}[h]
\begin{tabular}{c||c|c|c|c|c|c|c|}
{}
 & $\nu_{iR^{}}$ & $\psi_{aR}^{}$
 & $\Phi$ & $s_1^+$ & $\eta$ & $s_2^+$ \\
\hline%----------------------
\hline%----------------------
Spin $J$
 & $1/2$ & $1/2$
 & $0$ & $0$ & $0$ &  $0$ \\
\hline%----------------------
\hline%----------------------
%$\mathrm{SU(3)_C}$
% & 1 & 1
% & 1 & 1 & 1 & 1
%\\
%\hline%----------------------
$\mathrm{SU(2)_L}$
 & 1 & 1
 & 2 & 1 & 2 & 1
\\
\hline%----------------------
$\mathrm{U(1)_Y}$
 & 0 & 0
 & $3/2$ & 1 & $1/2$ & 1
\\
\hline%----------------------
$Z_2'$
 & $-$ & +
 & (+) & $-$ & + & +
\\
\hline%----------------------
$L$
 & 1 & 0
 & $-2$ & $-2$ & $-1$ & $-1$
\\
\hline%----------------------
\hline%----------------------
$Z_2$
 & + & $-$
 & + & + & $-$ & $-$
\\
\hline%----------------------
\end{tabular}
\caption{
The list of new fields in our model.
}
\label{table:particle list}
\end{table}

%(Yukawa interaction)

 In our model, there are three new Yukawa interactions as
\begin{align}
\label{eq:Yukawa_int}
\mathcal{L}_{\mathrm{Yukawa}}
=
 &\hspace{10pt}
 \left( Y_1 \right)_{\ell i}
 \di{ (\ell_R)^c } \hspace{1pt} \nu_{iR}^{} \hspace{1pt} s_1^+
 +
 \left( Y_2 \right)_{\ell a}
 \di{ (\ell_R)^c } \hspace{1pt} \psi_{aR}^{} \hspace{1pt} s_2^+
 +
 \left( Y_\eta \right)_{\ell a}
 \di{ L_\ell } \hspace{1pt} \eta^c \psi_{aR}^{}
 +
 \mathrm{h.c.}
\end{align}
The scalar potential is given by
\bal
\label{eq:potential}
V
=
 &\hspace{10pt}
 \mu_1^2 |\phi|^2
 + \mu_2^2 |\Phi|^2
 + \mu_3^2 |s_1^+|^2
 + \mu_4^2 |\eta|^2
 + \mu_5^2 |s_2^+|^2
\nonumber \\
 &
 +
   \left(
    \sigma_1
    \Phi^\dagger \phi s_1^+
    + \mathrm{h.c.}
   \right)
 +
   \left(
    \sigma_2
    \Phi^\dagger \eta \hspace{1pt} s_2^+
    + \mathrm{h.c.}
   \right)
 +
   \left(
    \sigma_3
    \phi^\dagger \eta^c s_2^+
    + \mathrm{h.c.}
   \right)
\nonumber \\
 &
 +
   \lambda_\phi
   |\phi|^4
 +
   \lambda_\Phi
   |\Phi|^4
 +
   \lambda_1
   |s_1^+|^4
 +
   \lambda_\eta
   |\eta|^4
 +
   \lambda_2
   |s_2^+|^4
\nonumber \\
 &
 +
   \lambda_{\phi\Phi}
   |\phi|^2 |\Phi|^2
 +
   \lambda'_{\phi\Phi}
   |\phi^\dagger \Phi|^2
 +
   \lambda_{\phi\eta}
   |\phi|^2| \eta|^2
 +
   \lambda'_{\phi\eta}
   |\phi^\dagger \eta|^2
 +
   \lambda_{\Phi\eta}
   |\Phi|^2 |\eta|^2
 +
   \lambda'_{\Phi\eta}
   |\Phi^\dagger\eta|^2
\nonumber \\
 &
 +
   \sum_{k = 1}^2
   \left\{
      \lambda_{\phi k}
      |\phi|^2 |s_k|^2
    +
      \lambda_{\Phi k}
      |\Phi|^2 |s_k|^2
    +
      \lambda_{\eta k}
      |\eta|^2 |s_k|^2
   \right\}
 +
   \lambda_{12}
   |s_1^+|^2 |s_2^+|^2
\nonumber \\
 &
 +
   \left(
    \xi_1
    \eta^\dagger \Phi \eta^\dagger \phi^c
    + \mathrm{h.c.}
   \right)
 +
   \left(
    \xi_2
    \Phi^\dagger \phi^c \left( s_2^+ \right)^2
    + \mathrm{h.c.}
   \right) .
\eal
 Notice that
$\sigma_1$ is the soft breaking parameter for $Z_2^\prime$.%
\footnote{
 If $\Phi$ is taken to be odd under $Z_2^\prime$,
the soft breaking parameter is $\sigma_2$.
 Therefore,
a product $\sigma_1 \sigma_2$ breaks $Z_2^\prime$
independently of the $Z_2^\prime$-parity of $\Phi$.
}
 There are five complex coupling constants%
~($\sigma_1$, $\sigma_2$, $\sigma_3$, $\xi_1$, and $\xi_2$),
and two CP-violating phases remain as physical parameters
after redefinitions of phases of fields.%
\footnote{
 If we take $\Phi$ as a $Z_2^\prime$-odd field,
terms of $\xi_1$ and $\xi_2$ are replaced with
$
 \xi_3
 \Phi^\dagger \eta^c \hspace{1pt} s_1^+ \hspace{1pt} s_2^+
$.
 Then, only one CP-violating phase is physical.
}
 In this article,
coupling constants in the scalar potential
are taken to be real, just for simplicity.

%(mass formulae for scalar particles)

 The SM Higgs doublet field $\phi$ does not mix with
the other scalar fields in our model.
 Thus, identically to the SM,
the field can be expressed as
$\phi = ( G^+, ( v + h +i G^0 )/\sqrt{2}\, )^T$,
where
$v$~($= \sqrt{-\mu_1^2/\lambda_\phi} = 246\,\GeV$)
is the vacuum expectation value.
 The real component $h$ corresponds to the SM Higgs boson,
whose mass is given by
$m_h^{} = \sqrt{2\lambda_\phi}\, v$.
 Nambu-Goldston bosons~($G^+$ and $G^0$)
are absorbed by the longitudinally polarized weak gauge bosons by the electroweak symmetry breaking.

 Fields $\Phi^{++}$ and $\eta^0$ are mass eigenstates.
 Their squared masses are given by
\bal
m^2_{\Phi^{++}}
&
=
  \mu_2^2
 +
  \frac{1}{2} \lambda_{\phi\Phi} v^2 ,
\\
%---------
m_{\eta^0}^2
&
=
  \mu_4^2
 +
  \frac{1}{2}
  \left(
   \lambda_{\phi\eta} + \lambda'_{\eta\phi}
  \right)
  v^2 .
\eal
 Mass eigenstates $\pi_1^+$ and $\pi_2^+$,
which are singly-charged and have $L = -2$,
are obtained by linear combinations of $\Phi^+$ and $s_1^+$ as
\bal
\begin{pmatrix}
 \pi_1^+ \\
 \pi_2^+ \\
\end{pmatrix}
=
 U_{\theta}
 \begin{pmatrix}
  \Phi^+ \\
  s_1^+ \\
 \end{pmatrix} ,
\quad
%-----------
U_\theta
=
 \begin{pmatrix}
  \cos\theta & \sin\theta \\
  -\sin\theta & \cos\theta \\
 \end{pmatrix} ,
\eal
where the mixing angle $\theta$ is defined as
\bal
\tan 2\theta
=
 \frac{ -2 \left( M^2_{\Phi s_1} \right)_{12} }
      { \left( M_{\Phi s_1}^2 \right)_{22}
 -
 \left( M_{\Phi s_1}^2 \right)_{11}} ,
\quad
%----------
M^2_{\Phi s_1}
=
 \begin{pmatrix}
   \mu_2^2
   + \frac{1}{2}
     \left( \lambda_{\phi\Phi} + \lambda'_{\phi\Phi} \right) v^2
  &
   \frac{ 1 }{ \sqrt{2} } \sigma_1 v \\
  %--------
   \frac{1}{ \sqrt{2} } \sigma_1 v
  &
   \mu_3^2
   + \frac{1}{2} \lambda_{\phi1} v^2 \\
 \end{pmatrix}  .
\eal
 Squared masses of $\pi_1^+$ and $\pi_2^+$ are given by
\bal
m_{\pi_1}^2
&=
 \frac{1}{\,2\,}
 \Biggl\{
   \left( M^2_{\Phi s_1} \right)_{11}
  +
   \left( M^2_{\Phi s_1} \right)_{22}
  +
   \sqrt{
     \left(
       \left( M^2_{\Phi s_1} \right)_{22}
      -
       \left( M^2_{\Phi s_1} \right)_{11}
     \right)^2
    +
     4 \left( M^2_{\Phi s_1} \right)^2_{12}
   }\,
 \Biggr\} ,
\\
%---------
m_{\pi_2}^2
&=
 \frac{1}{\,2\,}
 \Biggl\{
   \left( M^2_{\Phi s_1} \right)_{11}
  +
   \left( M^2_{\Phi s_1} \right)_{22}
  -
   \sqrt{
     \left(
       \left( M^2_{\Phi s_1} \right)_{22}
      -
       \left( M^2_{\Phi s_1} \right)_{11}
     \right)^2
    +
     4 \left( M^2_{\Phi s_1} \right)^2_{12}
   }\,
 \Biggr\} .
\eal

 Mass eigenstates $\omega_1^+$ and $\omega_2^+$,
which are $Z_2$-odd with $L = -1$,
are constructed by linear combinations of
$\eta^+$ and $s_2^+$ as follows:
\bal
&
\begin{pmatrix}
 \omega_1^+ \\
 \omega_2^+ \\
\end{pmatrix}
=
 U_{\chi}
 \begin{pmatrix}
  \eta^+ \\
  s_2^+ \\
 \end{pmatrix} ,
\quad
%-----------
U_\chi
=
 \begin{pmatrix}
  \cos\chi & \sin\chi \\
  -\sin\chi & \cos\chi \\
 \end{pmatrix} ,
\eal
where the mixing angle $\chi$ is defined as
\bal
\tan 2\chi
=
 \frac{ -2 \left( M^2_{\eta s_2} \right)_{12} }
      {
        \left( M_{\eta s_2}^2 \right)_{22}
        - \left(M_{\eta s_2}^2\right)_{11}
      } ,
\quad
%-----------
M^2_{\eta s_2}
=
 \begin{pmatrix}
   \mu_4^2
   + \frac{1}{2} \lambda_{\phi\eta} v^2
  &
   - \frac{1}{\sqrt{2}} \sigma_3 v \\
  %-------
   -\frac{1}{\sqrt{2}} \sigma_3 v
  &
   \mu_5^2
   + \frac{1}{2} \lambda_{\phi2} v^2 \\ 
 \end{pmatrix} .
\eal
 Squared masses of $\omega_1^+$ and $\omega_2^+$ are given by
\bal
m_{\omega_1}^2
&=
 \frac{1}{\,2\,}
 \Biggl\{
   \left( M^2_{\eta s_2} \right)_{11}
  +
   \left( M^2_{\eta s_2} \right)_{22}
  +
   \sqrt{
     \left(
       \left( M^2_{\eta s_2} \right)_{22}
      -
       \left( M^2_{\eta s_2} \right)_{11}
     \right)^2
    +
     4 \left( M^2_{\eta s_2} \right)^2_{12}
   }\,
 \Biggr\} ,
\\
%---------
m_{\omega_2}^2
&=
 \frac{1}{\,2\,}
 \Biggl\{
   \left( M^2_{\eta s_2} \right)_{11}
  +
   \left( M^2_{\eta s_2} \right)_{22}
  -
   \sqrt{
     \left(
       \left( M^2_{\eta s_2} \right)_{22}
      -
       \left( M^2_{\eta s_2} \right)_{11}
     \right)^2
    +
     4 \left( M^2_{\eta s_2} \right)^2_{12}
   }\,
 \Biggr\} .
\eal

\section{Neutrino Mass}
\label{sec:Neutrino Mass}
%(neutrino mass)

\begin{figure}[t]
\begin{center}
\includegraphics[width=100mm]{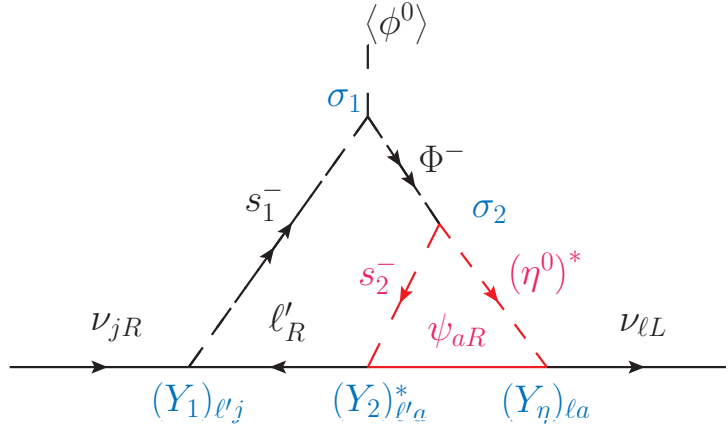}
\caption{
 The Feynman diagram to generate Dirac-type neutrino masses.
 Arrows denote flows of the conserved lepton number. Red colored lines represent those of $Z_2$-odd fields.
}
\label{fig:neutrino mass}
\end{center}
\end{figure}

 Mass terms
$(m_D^{})_{\ell i}\, \overline{\nu_{\ell L}^{}}\, \nu_{i R}^{}$
of Dirac neutrinos
are generated in our model
via two-loop diagrams in Fig.~\ref{fig:neutrino mass}.
%---------------------------
%---------------------------
 The Dirac neutrino mass matrix $(m_D^{})_{\ell i}$ is calculated as
\bal
\label{eq:neutrino_mass}
\left( m_\nu^{} \right)_{\ell i}
=&
 \frac{
       \Bigl( m^2_{\pi_2} - m^2_{\pi_1} \Bigr)
       \sigma_2
       \sin( 2\theta ) }
      { 2 }
 \sum_{\ell', a, k}
 \left( Y_\eta \right)_{\ell a}
 \left( Y_2 \right)^\ast_{\ell' a}
 \left( Y_1 \right)_{\ell' i}
 \left( U_\chi \right)_{k2}^2
 I_{\ell' a k} ,
\eal
where the coupling constant $\sigma_1$ in Fig.~\ref{fig:neutrino mass}
is replaced by using
$2 \sigma_1 \left<\phi^0\right>= \left( m^2_{\pi_1} - m^2_{\pi_2} \right) \sin( 2\theta )$.
 The explicit formula for the loop function $I_{\ell' a k}$
is given in Appendix~\ref{app:numass}.
 Notice that
$\sigma_2 \sin( 2\theta )$ softly breaks $Z_2^\prime$
that forbids $\di{L_\ell}\, \phi^c \nu_{iR}^{}$.

 Since we take the basis where $\nu_{i R}^{}$ are mass eigenstates,
the neutrino mass matrix $\left( m_\nu^{} \right)_{\ell i}$
is diagonalized as
\bal
m_\nu^{} = U_\MNS\, {\rm diag}(m_1, m_2, m_3) ,
\eal
where $m_i$~$(i = 1, 2, 3)$ denote masses of Dirac neutrinos.
The mixing matrix $U_\MNS$ is the
Maki-Nakagawa-Sakata matrix~\cite{Maki:1962mu},
which can be parameterized as
\bal
U_\MNS
&=
 \begin{pmatrix}
  1 & 0 & 0 \\
  0 & c_{23} & s_{23} \\
  0 & -s_{23} & c_{23} \\
 \end{pmatrix}
 \begin{pmatrix}
  c_{13} & 0 & s_{13}e^{-i\delta} \\
  0 & 1 & 0 \\
  -s_{13}e^{i\delta} & 0 & c_{13} \\
 \end{pmatrix}
 \begin{pmatrix}
  c_{12} & s_{12} & 0 \\
  -s_{12} & c_{12} & 0 \\
  0 & 0 & 1 \\
\end{pmatrix} ,
\eal
where $c_{ij} = \cos\theta_{ij}$ and
$s_{ij} = \sin\theta_{ij}$,
and $\delta$ is a CP-violating phase in the lepton sector.

\section{Lepton Flavor Violation}
\label{sec:Lepton Flavor Violation}
%---------------------------------------
\begin{figure}[t]
\begin{center}
\includegraphics[width=150mm]{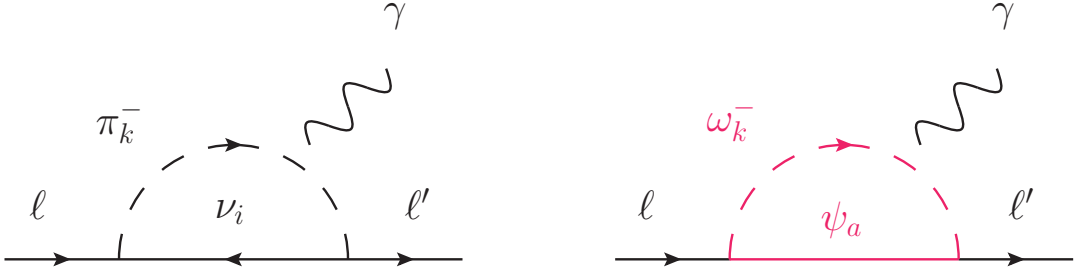}
\caption{Feynman diagrams for charged lepton LFV processes $\ell \to \ell^\prime \gamma$.}
\label{fig:LFV-lepton-decay}
\end{center}
\end{figure}
%---------------------------------------
Matrices $Y_1$, $Y_2$, and $Y_\eta$ are not diagonal
and cause LFV processes.
 Radiative decays of charged leptons, $\ell \to \ell^\prime \gamma$,
can be caused via the one-loop diagrams in Fig.~\ref{fig:LFV-lepton-decay}.
 Ignoreing $m_{\ell^\prime}^{}$,
branching ratios of these decays are expressed as
\bal
\frac{ \Br\left( \ell \to \ell' \gamma \right) }
     { \Br\left( \ell \to \ell' \nu_\ell^{} \overline{\nu_{\ell'}} \right) }
=
 \frac{3}{ 16\pi }
 \frac{ \alpha }{ G_F^2 m_\ell^4}
 \left( | A_R^{s_1} + A_R^\omega |^2 + | A_L^\omega |^2 \right)
,
\label{eq:LFV-lepton-decay}
\eal
where $G_F = 1.17 \times 10^{-5}\,\GeV^{-2}$ is the Fermi constant,
$\Br\left( \tau \to e \nu_\tau^{} \overline{\nu_e} \right)
\simeq
 0.178$,
$\Br\left( \tau \to \mu \nu_\tau^{} \overline{\nu_\mu} \right)
\simeq
 0.174$
and
$\Br\left( \mu \to e \nu_\mu^{} \overline{\nu_e} \right)
\simeq
 1$~\cite{Tanabashi:2018oca}.
 Formulae of $A_R^{s_1}$, $A_R^\omega$ and $A_L^\omega$
are presented in Appendix~\ref{app:ltolgamma}.
 $A_R^{s_1}$ corresponds to
the contribution from $s_1^+$ to $\ell \to \ell^\prime_R \gamma$.
 Contributions of $s_2^+$ and $\eta^+$
to $\ell \to \ell^\prime_R \gamma$ are given by $A_R^\omega$,
while $A_L^\omega$ is for their contributions
to $\ell \to \ell^\prime_L \gamma$.

\begin{figure}[t]
\begin{center}
\includegraphics[width=150mm]{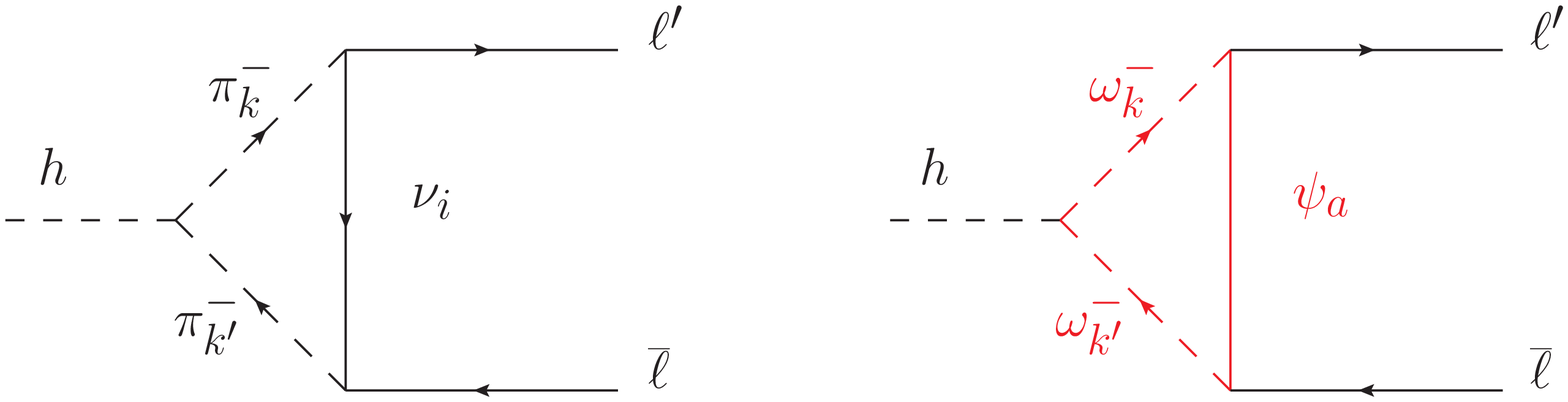}
\caption{
Diagrams for $h \to \ell \overline{\ell^\prime}$.}
\label{fig:LFV-Higgs-decay}
\end{center}
\end{figure}

 Scalar fields that contribute to $\ell \to \ell^\prime \gamma$
affect also $h \to \ell\ell^\prime$~($\ell \neq \ell^\prime$)
via diagrams in Fig.~\ref{fig:LFV-Higgs-decay}.
 Decay widths for $h \to \ell\ell^\prime$~($\ell \neq \ell^\prime$)
are given by
\bal
&
\Gamma\left( h \to \ell \ell^\prime \right)
=
  \Gamma\left( h \to \overline{\ell} \ell^\prime \right)
 +
  \Gamma( h \to \overline{\ell}{}^\prime \ell )
=\hspace{1pt}
 \frac{ m_h }{ 8\pi }
 \left( \frac{ 1 }{ 16\pi^2 } \right)^2
 \biggl(
   \bigl| B_R^{s_1} + B_R^\omega \bigr|^2
  +
   \bigl| B_L^\omega \bigr|^2
 \biggr) ,
\label{eq:HiggsLFV}
\eal
where we take $m_{\ell^\prime}^{} = 0$.
Formulae of $B_R^{s_1}$, $B_R^\omega$ and $B_L^\omega$
are shown in Appendix~\ref{app:htoll}.
 The contribution from $s_1^+$ is given by $B_R^{s_1}$,
while those from $s_2^+$ and $\eta^+$ are involved
in both of $B_R^\omega$ and $B_L^\omega$.
 The subscript $X$~$(= L, R)$ in these $B_X$'s indicates
the chirality of the lighter charged lepton $\ell^\prime_X$
in the final state.

\begin{figure}[t]
\begin{center}
\includegraphics[width=150truemm]{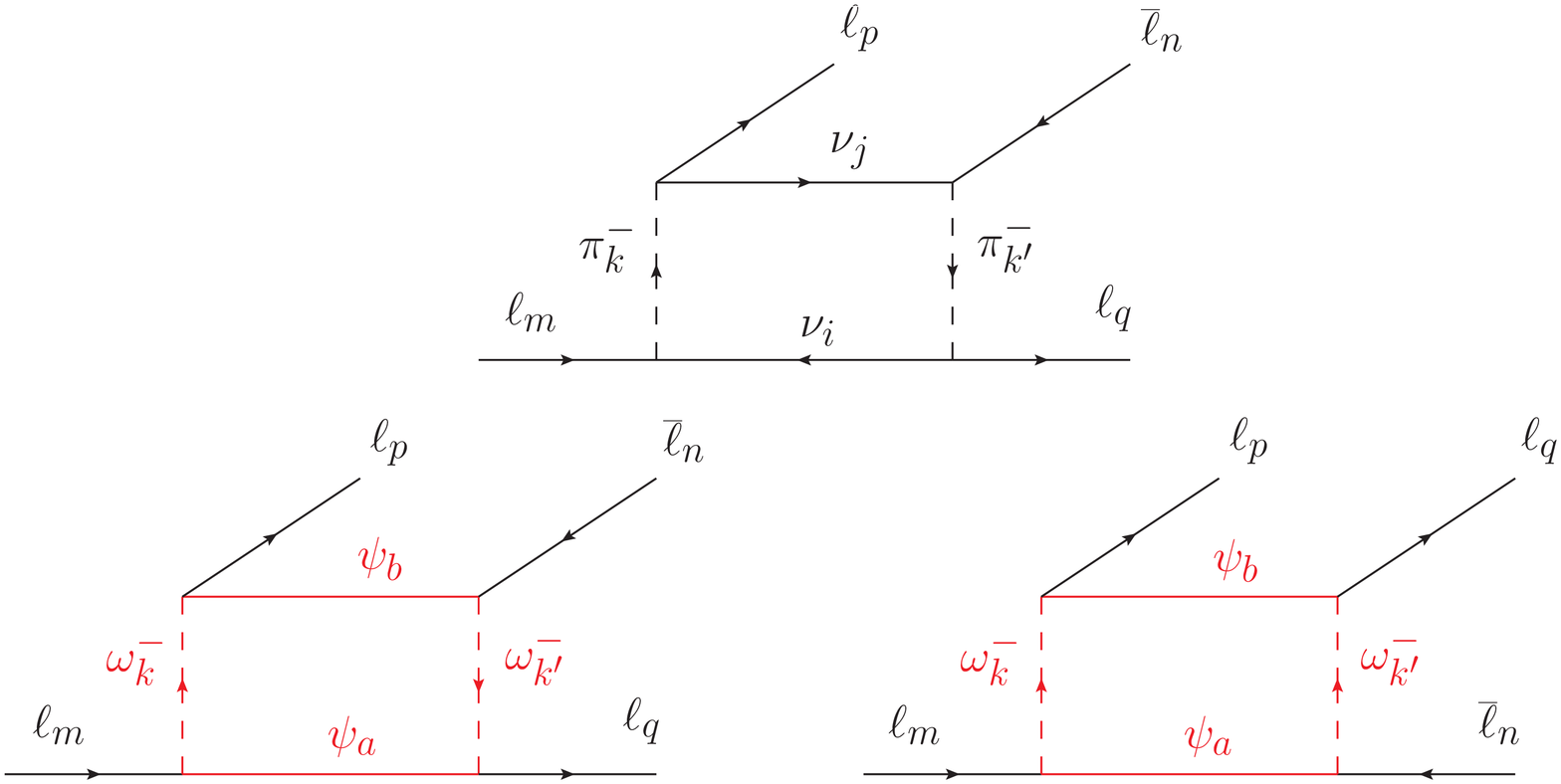}
\caption{
Feynman diagrams for $\ell_m \to \overline{\ell}_n \ell_p \ell_q$.
}
\label{fig:e_to_3e}
\end{center}
\end{figure}

 New scalar bosons in our model contribute also to
$\ell_m \to \overline{\ell}_n\, \ell_p\, \ell_q$%
~($m=2, 3$ and $n, p, q=1, 2$)
with new Yukawa interactions, where $\ell_1, \ell_2$ and $\ell_3$ corresponds to $e, \mu$ and $\tau$, respectively.
 Contributions from penguin diagrams can be ignored
because of the constraint from $\ell \to \ell^\prime \gamma$.
 However,
if some coupling constants of new Yukawa interactions are $O(1)$,
box diagrams in Fig.~\ref{fig:e_to_3e} should be considered.
 Branching ratios for
$\ell_m \to \overline{\ell}_n\, \ell_p\, \ell_q$ via the box diagrams
are given by
\bal
\frac{
 \Br
 \bigl(
  \ell_m \to  \overline{\ell}_n\, \ell_p\, \ell_q
 \bigr)
}{
 \Br
 \bigl(
  \ell_m \rightarrow \ell_p\, \nu_{\ell_m}^{} \overline{\nu}_{\ell_p}
 \bigr) 
}
&=
 \frac{ S }{ 64 G_F^2 }
 \left( \frac{1}{ 16\pi^2 } \right)^2
 \Biggl\{
   4
   \left(
     \Bigl|
       \left( C_{RRRR}^{s_1} \right)_{mnpq}
      +
       \left( C_{RRRR}^{s_2} \right)_{mnpq}
     \Bigr|^2
    +
     \Bigl|
      \left( C_{LLLL}^\eta \right)_{mnpq}
     \Bigr|^2
   \right)
\nonumber\\
&
{}+
   \left|
    \left( C_{LLRR}^\omega \right)_{mnpq}
   \right|^2
  +
   \left|
    \left( C_{LLRR}^\omega \right)_{mnqp}
   \right|^2
  -
   \text{Re}
   \left[
    \left( C_{LLRR}^\omega \right)_{mnpq}
    \left( C_{LLRR}^\omega \right)_{mnqp}^\ast
   \right]
\nonumber\\
&
{}+
   \left|
    \left( C_{RRLL}^\omega \right)_{mnpq}
   \right|^2
  +
   \left|
    \left( C_{RRLL}^\omega \right)_{mnqp}
   \right|^2
  -
   \text{Re}
   \left[
    \left( C_{RRLL}^\omega \right)_{mnpq}
    \left( C_{RRLL}^\omega \right)_{mnqp}^\ast
   \right]
\nonumber\\
&
{}+
   \left|
    \left( C_{LRRL}^\omega \right)_{mnpq}
   \right|^2
+
   \left|
    \left( C_{RLLR}^\omega \right)_{mnpq}
   \right|^2
  +
   \left|
    \left( C_{LRLR}^\omega \right)_{mnpq}
   \right|^2
  +
   \left|
    \left( C_{RLRL}^\omega \right)_{mnpq}
   \right|^2
 \Biggr\}
,
\label{eq:LFV_decay_via_box_diagram}
\eal
where $S = 1$~($2$) for $p=q$~($p \neq q$).
 The variable
$\left( C_{RRRR}^{s_1} \right)_{mnpq}$%
~( $\left( C_{RRRR}^{s_2} \right)_{mnpq}$ )
corresponds to the contribution from $s_1^+$~($s_2^+$)
in the first diagram~(the second and the third diagrams)
in Fig.~\ref{fig:e_to_3e}:
 the structure of chiralities
is $\ell_{mR} \to \overline{\ell_{nR}}\, \ell_{pR}\, \ell_{qR}$
because charged leptons
that have Yukawa interactions with $s_1^+$ and $s_2^+$
are only right-handed ones.
 The contribution from $\eta^+$ to
$\ell_{mL} \to \overline{\ell_{nL}}\, \ell_{pL}\, \ell_{qL}$
via the second and the third diagrams in Fig.~\ref{fig:e_to_3e}
is given by $\left( C_{LLLL}^\eta \right)_{mnpq}$.
 The other $\left( C^\omega \right)_{mnpq}$'s
arise due to the mixing between $s_2^+$ and $\eta^+$
in the second and the third diagrams in Fig.~\ref{fig:e_to_3e}.
 See Appendix~\ref{app:ltolll} for formulae of $\left( C \right)_{mnpq}$'s.
 Current constraints on the branching ratios for LFV processes%
~($\ell \to \ell^\prime \gamma$, $h \to \ell\ell^\prime$,
and $\ell_m \to \overline{\ell}_n\, \ell_p\, \ell_q$)
are summarized in Table~\ref{Table:Experimental Constrains of LFV}.

\begin{table}[t]
\begin{tabular}[t]{|c|c|}
\hline%------------------
Process & Upper limit
\\
\hline%------------------
\hline%------------------
 $\mu \to e \gamma$
  & $4.2 \times 10^{-13}\ \text{\cite{TheMEG:2016wtm}}$
\\
\hline%------------------
 $\tau \to e \gamma$
  & $3.3\times10^{-8}\ \text{\cite{Aubert:2009ag}}$
\\
\hline%------------------
 $\tau \to \mu \gamma$
  & $4.4 \times 10^{-8}\ \text{\cite{Aubert:2009ag}}$
\\
\hline%------------------
\end{tabular}
\hspace*{5mm}
\begin{tabular}[t]{|c|c|}
\hline%------------------
Process & Upper limit
\\
\hline%------------------
\hline%------------------
 $\mu \to \overline{e} ee$
  & $1.0 \times 10^{-12}\ \text{\cite{Bellgardt:1987du}}$
\\
\hline%------------------
 $\tau \to \overline{e} ee$
  & $2.7 \times 10^{-8}\ \text{\cite{Hayasaka:2010np}}$
\\
\hline%------------------
 $\tau \to \overline{\mu} e\mu$
  & $2.7 \times 10^{-8}\ \text{\cite{Hayasaka:2010np}}$
\\
\hline%------------------
 $\tau \to \overline{e} \mu\mu$
  & $1.7 \times 10^{-8}\ \text{\cite{Hayasaka:2010np}}$
\\
\hline%------------------
 $\tau \to \overline{e} e\mu$
  & $1.8 \times 10^{-8}\ \text{\cite{Hayasaka:2010np}}$
\\
\hline%------------------
 $\tau \to \overline{\mu} ee$
  & $1.5 \times 10^{-8}\ \text{\cite{Hayasaka:2010np}}$
\\
\hline%------------------
 $\tau \to \overline{\mu} \mu\mu$
  & $2.1 \times 10^{-8}\ \text{\cite{Hayasaka:2010np}}$
\\
\hline%------------------
\end{tabular}
\hspace*{5mm}
\begin{tabular}[t]{|c|c|}
\hline%------------------
Process & Upper limit
\\
\hline%------------------
\hline%------------------
 $h \to \mu e $
  & $3.5 \times 10^{-4}\ \text{\cite{Khachatryan:2016rke}}$
\\
\hline%------------------
 $h \to \tau e$
  & $6.1 \times 10^{-3}\ \text{\cite{Sirunyan:2017xzt}}$
\\
\hline%------------------
 $h \to \mu\tau$
  & $2.5\times 10^{-3}\ \text{\cite{Sirunyan:2017xzt}}$
\\
\hline%------------------
\end{tabular}
\caption{
 Current experimental constrains on branching ratios of LFV processes.
}
\label{Table:Experimental Constrains of LFV}
\end{table}

%-------------------------
%-------------------------

\section{Dark Matter}
\label{sec:Dark Matter}

 In our model,
dark matter candidates are
the lightest of fermions $\psi_a$ and a scalar $\eta^0$,
which are neutral $Z_2$-odd particles.
 Notice that
$\eta^0$ from a doublet field is a complex scalar
with the lepton number $L = -1$.
 In other words,
there is no mass spritting between
CP-even and odd parts of $\eta^0$.
According to Ref.~\cite{Escudero:2016gzx}, the scenario where the dark matter is such a complex scalar is stringently constrained from direct search experiments because it interacts with nuclei at tree level. Therefore, we consider the case where the dark matter is the lightest one of gauge singlet Majorana fermions $\psi_a$.

\begin{figure}[t]
\begin{center}
\includegraphics[width=150mm]{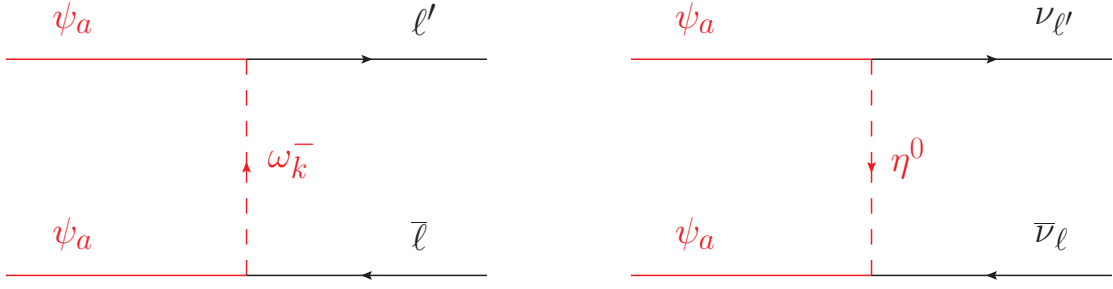}
\caption{
 Diagrams that give leading contributions to the DM relic abundance.
}
\label{fig:relic_abundance}
\end{center}
\end{figure}

 The dark matter candidate $\psi_a$ can be annihilated
via tree-level diagrams shown in Fig.~\ref{fig:relic_abundance}.
 The thermal averages $\left< \sigma v_\mathrm{rel} \right>$,
where $\sigma$ is annihilation cross section of $\psi_a$
and $v_\mathrm{rel}$ denotes the relative velocity of the initial particles,
is given by a sum of two processes as
$\left< \sigma v_\mathrm{rel} \right>
=
  \left< \sigma_\ell^{} v_\mathrm{rel} \right>
 +
  \left< \sigma_\nu v_\mathrm{rel} \right>$.
 Thermal avalages $\left< \sigma_\ell^{} v_\mathrm{rel} \right>$
and $\left< \sigma_\nu v_\mathrm{rel} \right>$
correspond to the effects of left and right diagrams
in Fig.~\ref{fig:relic_abundance}, respectively.
 Formulae of $\left< \sigma_\ell^{} v_\mathrm{rel} \right>$
and $\left< \sigma_\nu v_\mathrm{rel} \right>$
are shown in Appendix~\ref{app:DM}.
 Notice that
the $s$-wave annihilation is only involved
in $\left< \sigma_\ell^{} v_\mathrm{rel} \right>$
with a mixing $\chi$.

 For the case where the elements of $Y_\eta$ and the mixing angle $\chi$ are negligible%
~(we take such a benchmark scenario in the next section),
the dominant contribution to $\left< \sigma v_\mathrm{rel} \right>$
comes from the mediation of $s_2^+$~($\simeq \omega_2^+$)
in the left diagram in Fig.~\ref{fig:relic_abundance}.
 Then,
$\left< \sigma v_\mathrm{rel} \right>$ is approximately calculated as
\bal
\left< \sigma v_\mathrm{rel} \right>
\simeq
 \frac{ 1 }{ 8\pi }
 \bigl( Y_2^\dagger Y_2 \bigr)_{aa}^2
 \frac{
       M_{\psi_a}^2
       \bigl( M_{\psi_a}^4 + m_{\omega_2}^4 \bigr)
      }
      {
       \bigl( M_{\psi_a}^2 + m_{\omega_2}^2 \bigr)^4
      }\,
 \frac{ 1 }{\,x\,} ,
\label{eq:sigmav}
\eal
where
$x = m_{\psi_a}^{}/T$ at the temperature $T$.
In Appendix E, $\left<\sigma v_\mathrm{rel}\right>$ for more general case is presented.
 The relic abundance of $\psi_a$
with the $p$-wave annihilation is calculated as
\bal
\Omega_{\psi_a} h^2
=
  1.04 \times 10^{9} \times 2 \times x_f^2
  \frac{ \sqrt{g_*} }{ g_{*s} }
  \left\{
   \frac{ \GeV^{-1} }
        {
         m_\Pl^{}
         \left< \sigma v_\mathrm{rel} \right>|_{x=1}
        }
  \right\} ,
\label{eq:relic_abundance}
\eal
where
$m_\Pl^{} = 1.2 \times 10^{19}\,\GeV$ stands for the Planck mass,
and $g_* = 106.75$~($g_{*S} = 106.75$) is the effective degree of freedom
for energy~(entropy) density in the era of the freeze out of the dark matter%
~\cite{Kolb:1990vq}, and
$x_f$ is defined by
\bal
x_f
=&
  \ln
  \bigl[
   0.038 \times 2 ( g_\psi^{}/\sqrt{g_*} )
   m_\Pl^{} M_{\psi_a}
   \left< \sigma v_\mathrm{rel} \right>|_{x=1}
  \bigr]
\nonumber\\
&
 -
  \frac{3}{2}
  \ln
  \biggl\{
   \ln
   \bigl[
    0.038 \times 2 (g_\psi^{}/\sqrt{g_*})
    m_\Pl^{} M_{\psi_a}
    \left< \sigma v_\mathrm{rel} \right>|_{x=1}
   \bigr]
  \biggr\} ,
\label{eq:xf}
\eal
where
$g_\psi^{} = 2$ is the degree of freedom of $\psi_a$.

%--------------------------
%--------------------------

\section{Benchmark Scenarios and Numerical Evaluation}
\label{sec:Benchmark Scenarios and Numerical Evaluation}

 We here consider the possibility
that $h \to \mu\tau$ is enhanced
in comparison with LFV decays of charged leptons.
First,
we take the following benchmark scenario
for $m_1 < m_3$~(the normal ordering case of neutrino masses):
\begin{align}
\label{eq:benchmark_normal}
&
 \sigma_2 = 50\,\GeV, \hspace{5pt}
%--------
 \sin{2\theta} \simeq -3.38\times 10^{-2}, \hspace{5pt}
%--------
 \sin{2\chi} \simeq 4.99 \times10^{-7},
\nonumber\\
%----------
&
 \lambda_{\phi1} = 1.0, \hspace{5pt}
%----------
 \lambda_{\phi2} = -1.0, \hspace{5pt}
%----------
 \lambda_{\phi\eta} = -1.0, \hspace{5pt}
%----------
 \lambda_{\phi\Phi} = \lambda'_{\phi\Phi} = 1.0,
\nonumber\\
%----------
&
 M_{\psi_1} = 97.6\,\GeV, \hspace{10pt}
%----------
 M_{\psi_2} = 150\,\GeV, \hspace{10pt}
%----------
 M_{\psi_3} = 200\,\GeV,
\nonumber\\
%----------
&
 m_{\pi_1}^{} = 520\,\GeV, \hspace{10pt}
%----------
 m_{\pi_2}^{} = 510\,\GeV,
\nonumber\\
%----------
&
 m_{\omega_1}^{} \bigl( m_{\eta^+}^{} \bigr) = 1000\,\GeV, \hspace{10pt}
%----------
 m_{\omega_2}^{} \bigl( m_{s_2}^{} \bigr) = 550\,\GeV, \hspace{10pt}
%----------
 m_{\eta^0}^{} = 1000\,\GeV ,
\nonumber\\
%----------
&
Y_1
=
 \begin{pmatrix}
  10^{-4} & 10^{-4} & 0.10 \\
  0.80 & 0.80 & 10^{-4} \\
  1.5 & 1.0 & 10^{-4}
 \end{pmatrix} ,
\nonumber\\
%----------
&
Y_2
=
 \begin{pmatrix}
  10^{-4} & 10^{-4} & 0.10 \\
  -1.5 & -0.40 & 10^{-4} \\
  1.5 & 1.0 & 10^{-4}
 \end{pmatrix}, \hspace{10pt}
%----------
Y_\eta
\simeq
 \begin{pmatrix}
6.60\times10^{-4} & -4.56\times10^{-4} & 4.22\times10^{-3} \\
1.98\times10^{-4} & -2.65\times 10^{-4} & 1.85\times10^{-2} \\
-2.88\times10^{-4} & 2.98\times10^{-4} & 2.19\times10^{-2} \end{pmatrix} .
\end{align}
 The small value of the mixing angle $\chi$
implies $\omega_1^+ \simeq \eta^+$
and $\omega_2^+ \simeq s_2^+$.
 Since $\pi_k^+$ and  $\omega_k^+$
have Yukawa interactions only with leptons,
their masses are constrained
by the slepton searches in the context of supersymmetric models at the LHC,
which give about $500\,\GeV$ as the lower bound~\cite{Aaboud:2018jiw}.

 The generated neutrino mass matrix
results in the following values,
which are consistent with the current constraint
from neutrino oscillation experiments~\cite{Tanabashi:2018oca}:
\bal
&
 \sin^2{\theta_{12}} = 0.307 , \hspace{10pt}
%------------
 \sin^2{\theta_{13}} = 2.12 \times10^{-2} , \hspace{10pt}
%------------
 \sin^2{\theta_{23}} = 0.417 ,
\\
%------------
&
 \Delta m^2_{21} = 7.53 \times10^{-5}\,\eV^2 ,
\\
%------------
&
 \Delta m^2_{32} = 2.51 \times 10^{-3}\,\eV^2 ,
\\
%------------
&
 m_1 =0.048 \,\eV ,
\\
%------------
&
 \delta = 0,
\eal
 where $\Delta m_{ij}^2=m_i^2-m_j^2$. The values of $m_1$ is also consistent with
$\sum_i m_i < 0.26\,\eV$
that is given by cosmological observations~\cite{Loureiro:2018pdz},
although $m_1$ is not constrained by the oscillation data.

 In Table~\ref{tab:LFV-NH}, we show branching ratios for the LFV processes $\ell \to \ell' \gamma$, $\ell_m \to \overline{\ell}_n \ell_p \ell_q$ and $h\to \ell \ell'$ in our benchmark scenario given in Eq.~\eqref{eq:benchmark_normal}. They satisfy the constraints from the current data in Table~\ref{Table:Experimental Constrains of LFV}.
 Since the elements of $Y_\eta$
are rather small as seen in Eq.~\eqref{eq:benchmark_normal},
the contribution from $\eta^+$ to $\ell \to \ell^\prime \gamma$\ 
($A_L^\omega$ in Eq.~\eqref{eq:LFV-lepton-decay}) is negligible.
 Then,
values of $\BR(\ell \to \ell^\prime \gamma)$ in our benchmark scenario
are suppressed due to the cancellation of $A_R^{s1}$ and $A_R^\omega$,
which are contributions from $s_1^+$ and $s_2^+$, respectively.
 This is an interesting utilization
of scalar bosons~($s_1^+$ and $s_2^+$)
that are originally introduced for generating neutrino masses.
 On the other hand,
the contribution from $\eta^+$ to $h \to \ell\ell^\prime$%
~($B_L^\omega$ in Eq.~\eqref{eq:HiggsLFV})
is also negligible due to small values of components of $Y_\eta$.
 Even though
contributions from $s_1^+$ and $s_2^+$ to $\ell \to \ell^\prime \gamma$
are destructive with each other,
their contributions to $h \to \ell\ell^\prime$%
~($B_R^{s1}$ and $B_R^\omega$ in Eq.~\eqref{eq:HiggsLFV})
are not necessarily cancelled with each other
because of the sign flip by using coupling constants in the scalar sector, 
$\Lambda^\pi_{22}$ and $\Lambda^\omega_{22}$ in Appendix~\ref{app:htoll}.~\footnote{Notice that other $\Lambda^\pi${}'s and $\Lambda^\omega${}'s do not contribute to the cancellation dominantly because $\theta$ and $\chi$ are small in the benchmark scenario.}
 In our benchmark scenario,
$\BR( h \to \mu\tau )$ is indeed
much larger than $\BR( \tau \to \mu \gamma )$.
 This hierarchy is what expected in Ref.~\cite{Aoki:2016wyl},
and our calculation explicitly shows that the expectation is correct.
\begin{table}[t]
\begin{tabular}[t]{|c|c|}
\hline%------------------
Process & Numerical result
\\
\hline%------------------
\hline%------------------
 $\mu \to e \gamma$
  & $3.36 \times 10^{-16}$
\\
\hline%------------------
 $\tau \to e \gamma$
  & $1.25 \times 10^{-14}$
\\
\hline%------------------
 $\tau \to \mu \gamma$
  & $1.18 \times 10^{-9}$
\\
\hline%------------------
\end{tabular}
\hspace*{5mm}
\begin{tabular}[t]{|c|c|}
\hline%------------------
Process & Numerical result
\\
\hline%------------------
\hline%------------------
 $\mu \to \overline{e} ee$
  & $2.73 \times 10^{-19}$
\\
\hline%------------------
 $\tau \to \overline{e} ee$
  & $6.23 \times 10^{-19}$
\\
\hline%------------------
 $\tau \to \overline{\mu} e\mu$
  & $2.98 \times 10^{-14}$
\\
\hline%------------------
 $\tau \to \overline{e} \mu\mu$
  & $1.42 \times 10^{-14}$
\\
\hline%------------------
 $\tau \to \overline{e} e\mu$
  & $3.08 \times 10^{-12}$
\\
\hline%------------------
 $\tau \to \overline{\mu} ee$
  & $1.57 \times 10^{-13}$
\\
\hline%------------------
 $\tau \to \overline{\mu} \mu\mu$
  & $4.50 \times 10^{-10}$
\\
\hline%------------------
\end{tabular}
\hspace*{5mm}
\begin{tabular}[t]{|c|c|}
\hline%------------------
Process & Numerical result
\\
\hline%------------------
\hline%------------------
 $h \to \mu e$
  & $2.10 \times 10^{-18}$
\\
\hline%------------------
 $h \to \tau e$
  & $1.57 \times 10^{-17}$
\\
\hline%------------------
 $h \to \mu \tau$
  & $1.04 \times 10^{-7}$
\\
\hline%------------------
\end{tabular}
\caption{
Numerical results for the LFV branching ratios in the benchmark scenario
for the normal ordering case.
}
\label{tab:LFV-NH}
\end{table}

\begin{figure}[t]
\centering
\includegraphics[width=70mm]{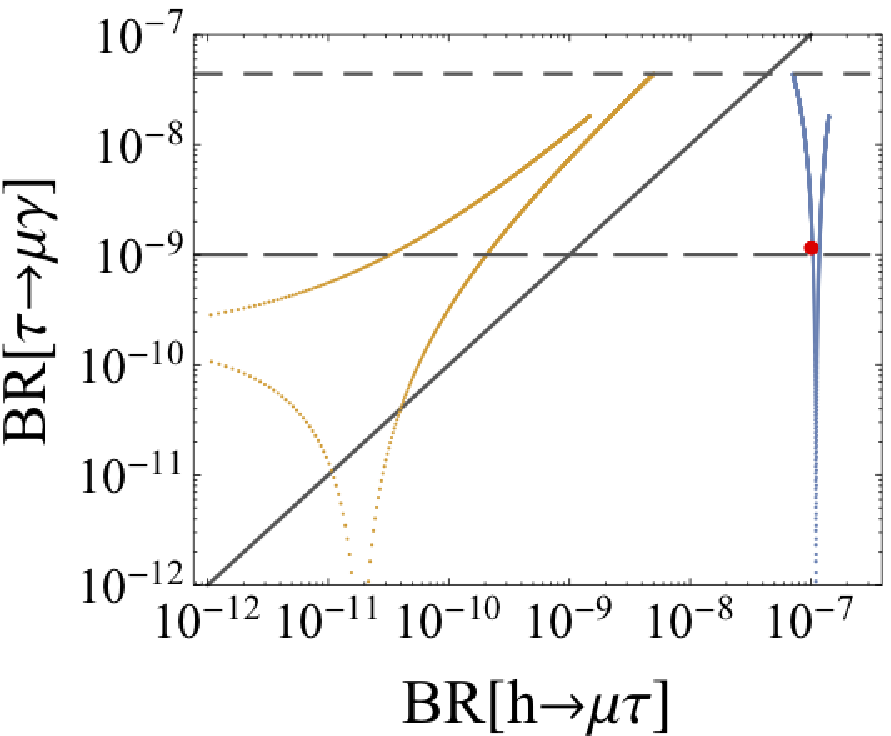}
\hspace*{5mm}
\includegraphics[width=70mm]{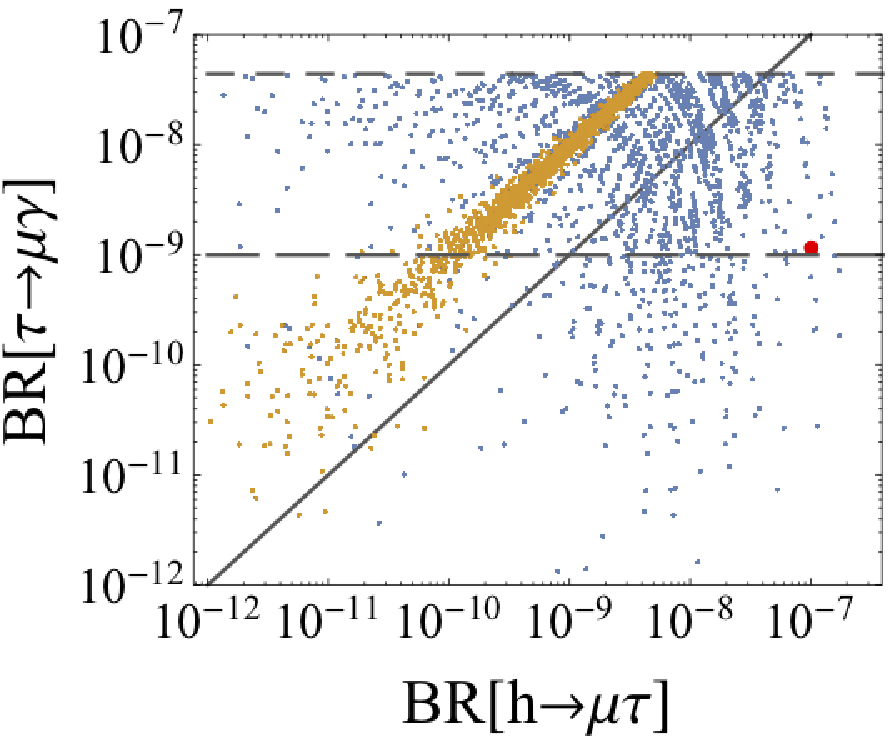}
\caption{Plots of the branching ratio for $\tau \to \mu \gamma$ versus that for $h \to \mu \tau$. }
\label{fig:Higgs_mu_tau_vs_Tau_mu_gamma}
\end{figure}

In Fig.~\ref{fig:Higgs_mu_tau_vs_Tau_mu_gamma}, we show plots of the branching ratio for $\tau\to\mu\gamma$ versus that for $h\to\mu\tau$. In the left one, we change only the values of $\left(Y_1\right)_{\mu2}$ between $-1.5$ and $1.5$. In the right one, we assume that the form of the matrices $Y_1$ and $Y_2$ is 
\bal
Y_k = 
\left(
\begin{array}{ccc}
10^{-4} & 10^{-4} & 0.10 \\ 
\left(Y_k\right)_{\mu 1} & \left(Y_k\right)_{\mu 2} & 10^{-4} \\
\left(Y_k\right)_{\tau 1} & \left(Y_k\right)_{\tau 2} & 10^{-4} \\
\end{array}
\right),
\label{eq:Assumption_for_plot}
\eal
where $k=1,2$, and then we vary eight unfixed parameters between $-1.5$ and $1.5$. 
The orange points are predictions with same sign $\lambda${}'s, $\lambda_{\phi 1}=\lambda_{\phi 2}=\lambda_{\phi\eta}=\lambda_{\phi \Phi}=\lambda_{\phi \Phi}'=1.0$. 
The blue points are ones with opposite sign $\lambda${}'s, $\lambda_{\phi 1}=\lambda_{\phi \Phi}=\lambda_{\phi \Phi}'=1.0$ and $\lambda_{\phi 2}=\lambda_{\phi\eta}=-1.0$,
as in the benchmark scenario. 
In both of the plots, values of fixed parameters are taken to be the same with those of the benchmark scenario. 
Two branching ratios are equal on the solid line in the figures. 
The upper dashed line is the current upper limit for $\BR(\tau \to \mu \gamma)$, $4.4 \times 10^{-8}$, and the lower one is the expected upper limit, $1.0 \times 10^{-9}$, from the Belle-II experiment \cite{Kou:2018nap} with the integrated luminosity $50 ~\mathrm{ab^{-1}}$. 
In the case with same sign $\lambda${}'s, the correlation between branching ratios is almost linear,
and $\BR(\tau \to \mu \gamma)$ is larger than $\BR(h \to \mu \tau)$ in most of the orange points. 
In the case with opposite sign $\lambda${}'s, on the other hand, $\BR(h \to \mu \tau)$ are significantly larger than $\BR(\tau \to \mu \gamma)$ in some of the blue points. This is just what we anticipated. The red point represents the result in the benchmark scenario.

\begin{figure}[t]
\centering
\includegraphics[width=70mm]{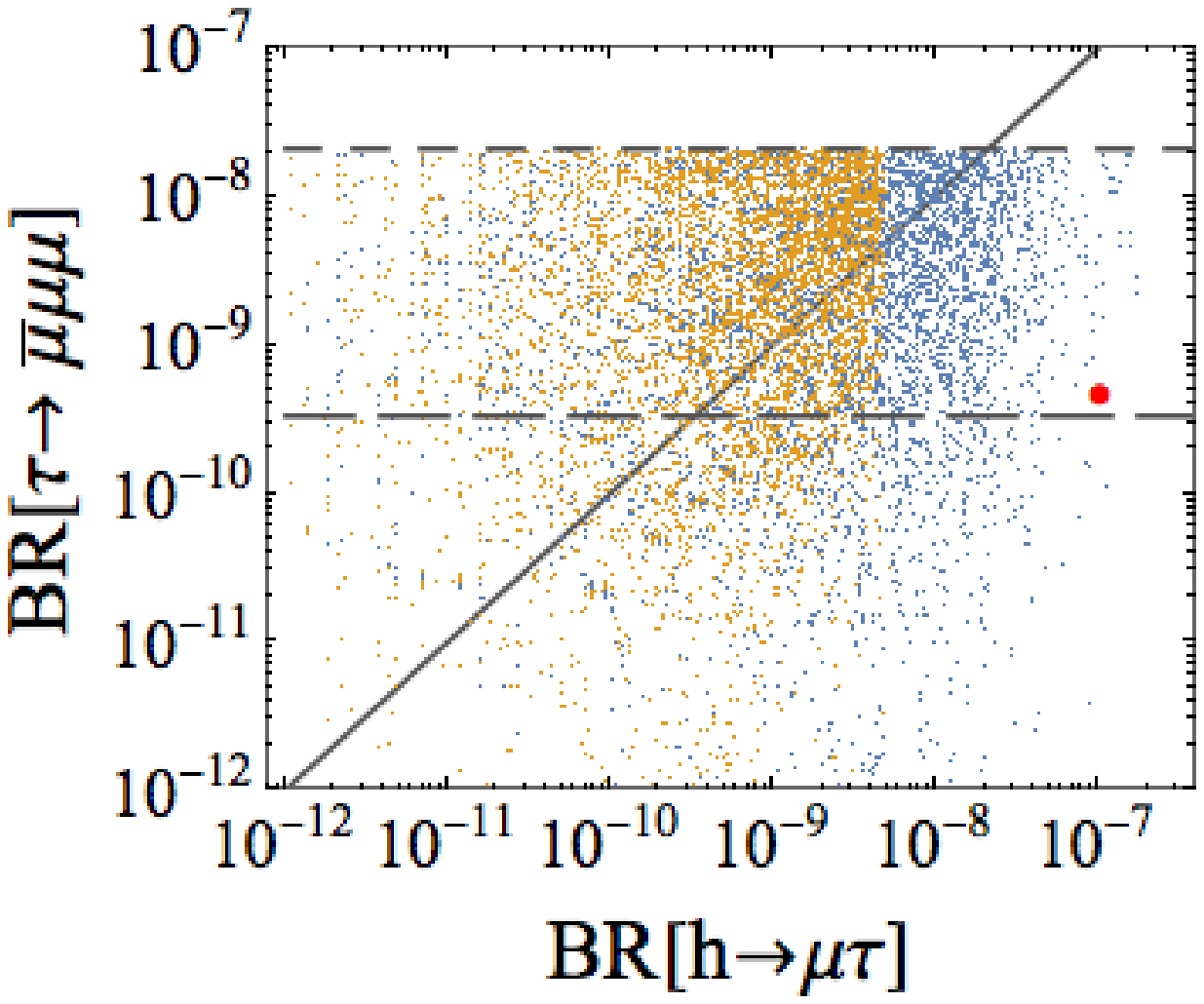}
\caption{
 The plot of the branching ratio for $\tau \to \overline{\mu}\mu\mu$ versus that for $h \to \mu \tau$.
}
\label{fig:Higgs_mu_tau_vs_Tau_3mu}
\end{figure}

In Fig.~\ref{fig:Higgs_mu_tau_vs_Tau_3mu}, we show the plot for $\BR(\tau \to  \overline{\mu}\mu\mu)$ versus $\BR(h\to\mu\tau)$ under the same assumptions as in the right one of Fig.~\ref{fig:Higgs_mu_tau_vs_Tau_mu_gamma}.
The upper dashed line is the current upper limit for $\BR(\tau \to  \overline{\mu}\mu\mu)$,
$2.1\times 10^{-8}$,
and the lower one is the expected upper limit, $3.3 \times 10^{-10}$, from the Belle-II experiment~\cite{Kou:2018nap} with the integrated luminosity $50~\mathrm{ab^{-1}}$.
We cannot find any correlation between the branching ratios, because these processes are given by different kind of Feynman diagrams.

Although
$\BR( h \to \mu\tau ) = 1.04 \times 10^{-7}$
is about $10^2$ times larger than
our prediction on
$\BR( \tau \to \mu \gamma )$,
the value is rather below the expected sensitivities,
$O(10^{-4})$ at HL-LHC~\cite{Calibbi:2017uvl,Banerjee:2016foh}
and $O(10^{-5})$ at ILC250~\cite{Chakraborty:2016gff}.
 We can in principle enhance $\BR( h \to \mu\tau )$ further
by taking larger values%
\footnote{
 It is difficult to take lighter masses of $s_1^+$ and $s_2^+$
because of constraint from the slepton searches.
}%
of $Y_1$, $Y_2$, and $\lambda$'s,
although we should worry about unitality bounds.
On the other hand,
the values for $\BR( \tau \to \mu \gamma)$ and $\BR( \tau \to \bar{\mu} \mu \mu)$ in our benchmark scenario
are close to the sensitivity in Belle II experiment~\cite{Kou:2018nap},
and then the scenario might be tested. In the case that $\tau \to \mu \gamma$ or $\tau \to \overline{\mu} \mu\mu$ is observed, we can distinguish our benchmark scenario from other models for tiny neutrino masses by the searches for the signal of $h \to \mu\tau$.

\begin{figure}[t]
\centering
\includegraphics[width=70mm]{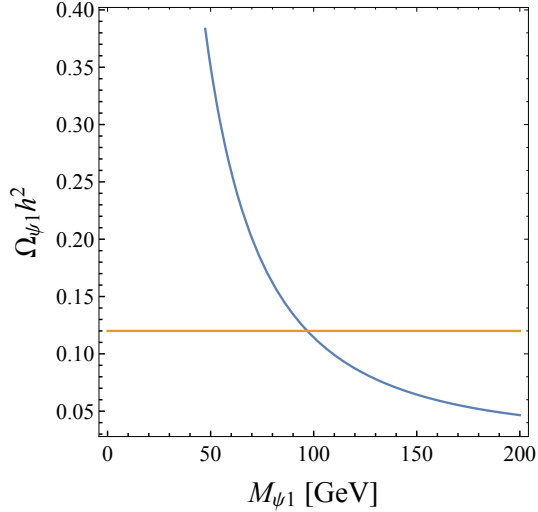}
\caption{
 Mass dependence of the DM relic abundance in the benchmark scenario
for the normal ordering case of neutrino masses.
}
\label{fig:psi_1_relic}
\end{figure}

 The dark matter in the benchmark scenario is
the lightest $Z_2$-odd Majorana fermion $\psi_1$.
 The density of the thermal relic abundance $\Omega_{\psi_1} h^2$ can be evaluated 
with Eqs.~\eqref{eq:sigmav}-\eqref{eq:xf},
which are valid for the case where $Y_\eta$ and $\chi$ are negligible.
The Planck experiment shows that
$\Omega_\mathrm{DM} h^2 = 0.1200 \pm 0.0012$~\cite{Aghanim:2018eyx}.
 In Fig.~\ref{fig:psi_1_relic},
we show $\Omega_{\psi_1} h^2$ as a function of
the dark matter mass $M_{\psi_1}$,
where $Y_2$ and $m_\omega^{}$ are fixed
to the values of the benchmark scenario.
 The blue curve is the mass dependence in our model,
and the horizontal line shows the observed value.
 It is clear that
the appropriate value of $\Omega_{\psi_1} h^2$
is obtained for
$M_{\psi_1} = 97.6\,\GeV$
in the benchmark scenario.

There is no tree-level contribution to the dark matter scattering off nuclei, because $\psi_a$ are gauge singlet fermions. 
The scattering occurs at one-loop level via three penguin diagrams with $\omega_1,\ \omega_2$ and $\eta^0$ in the loop. In our benchmark scenario, the elements of the matrix $Y_\eta$ are typically smaller than those of $Y_2$, so that we consider only the contribution from the diagram with $\omega_2$ in the loop. 
In Ref.~\cite{Herrero-Garcia:2018koq}, the authors studied in detail the gauge singlet Majorana dark matter which is coupled to a dark scalar and charged leptons. 
They also considered the scenario where the dark matter has no interaction with electrons, which is similar to our benchmark scenario. 
They gave the constraint from the direct searches with the combined data from XENON1T~\cite{Aprile:2017iyp}, PandaX~\cite{Cui:2017nnn} and LUX~\cite{Akerib:2016vxi}. 
From their results, we can estimate that the upper limits on $(Y_2)_{\mu1}$ and $(Y_2)_{\tau1}$ in our benchmark scenario are both about $3.2$.
Since $\left(Y_2\right)_{\mu 1}$ and $\left(Y_2\right)_{\tau 1}$ in Eq.~(\ref{eq:benchmark_normal}) are below this upper limit, the dark matter in our benchmark scenario satisfies the constraint from the current direct detection experiments.

Next, we consider the benchmark scenario for the inverted ordering case ($m_{3} < m_{1}$). The difference from the normal ordering case appears on $Y_\eta$, and we here take

\begin{align*}
Y_\eta\simeq
\begin{pmatrix}
1.17\times10^{-3} & -8.08\times10^{-4} & 4.22\times10^{-3} \\
3.41\times10^{-4} & -4.59\times 10^{-4} & 1.86\times10^{-2} \\
-5.11\times10^{-4} & 5.34\times10^{-4} & 2.18\times10^{-2}
\end{pmatrix}
.\\
\end{align*}
All the other parameters are taken to be the same with those in Eq. (\ref{eq:benchmark_normal}).

The neutrino mass matrix generated at two loop gives the following values, which are consistent with the current constraint from neutrino oscillation experiments~\cite{Tanabashi:2018oca},

\bal
&
 \sin^2{\theta_{12}} = 0.307 , \hspace{10pt}
%------------
 \sin^2{\theta_{13}} = 2.12 \times10^{-2} , \hspace{10pt}
%------------
 \sin^2{\theta_{23}} = 0.421 ,
\\
%------------
&
 \Delta m^2_{21} = 7.53 \times10^{-5}\,\eV^2 ,
\\
%------------
&
 \Delta m^2_{32}  = -2.56 \times 10^{-3}\,\eV^2 ,
\\
%------------
&
 m_3 =0.07 \,\eV,
\\
%------------
&
 \delta = 0 .
\eal
 The value of $m_3$ satisfies the condition from the Planck ovservation, $\sum_i m_{i}<0.26\ \mathrm{eV}$ ~\cite{Loureiro:2018pdz}.

Branching ratios for the LFV processes in this scenario are listed in Table~\ref{Numerical result in benchmark scenario (IH)}.
\begin{table}[t]
\begin{tabular}[t]{|c|c|}
\hline%------------------
Process & Numerical result
\\
\hline%------------------
\hline%------------------
 $\mu \to e \gamma$
  & $3.34 \times 10^{-16} $
\\
\hline%------------------
 $\tau \to e\gamma$
  & $1.25 \times 10^{-14}$
\\
\hline%------------------
 $\tau \to \mu \gamma$
  & $1.21 \times 10^{-9}$
\\
\hline%------------------
\end{tabular}
\hspace*{5mm}
\begin{tabular}[t]{|c|c|}
\hline%------------------
Process & Numerical result
\\
\hline%------------------
\hline%------------------
 $\mu \to \overline{e} ee$
  & $2.74 \times 10^{-19}$
\\
\hline%------------------
 $\tau \to \overline{e} ee$
  & $6.24 \times 10^{-19}$
\\
\hline%------------------
 $\tau \to \overline{\mu} e\mu$
  & $2.98 \times 10^{-14}$
\\
\hline%------------------
 $\tau \to \overline{e} \mu \mu$
  & $1.42 \times 10^{-14}$
\\
\hline%------------------
 $\tau \to \overline{e} e\mu$
  & $3.08 \times 10^{-12}$
\\
\hline%------------------
 $\tau \to \overline{\mu} ee$
  & $1.57 \times 10^{-13}$
\\
\hline%------------------
 $\tau \to \overline{\mu} \mu\mu$
  & $4.50 \times 10^{-10}$
\\
\hline%------------------
\end{tabular}
\hspace*{5mm}
\begin{tabular}[t]{|c|c|}
\hline%------------------
Process & Numerical result
\\
\hline%------------------
\hline%------------------
 $h \to \mu e$
  & $2.10 \times 10^{-18}$
\\
\hline%------------------
 $h \to \tau e$
  & $1.56 \times 10^{-17}$
\\
\hline%------------------
 $h \to \mu \tau$
  & $1.04 \times 10^{-7}$
\\
\hline%------------------
\end{tabular}
\caption{Numerical results for the LFV branching ratios in the benchmark scenario for the inverted ordering case of neutrino masses.}
\label{Numerical result in benchmark scenario (IH)}
\end{table}
Most of all branching ratios are the same as those in the scenario in Eq.(\ref{eq:benchmark_normal}), because the elements of $Y_\eta$ are typically smaller than those of $Y_1$ and $Y_2$ in the both scenarios. 
$\BR(h\to \mu\tau)$ is about $10^2$ times larger than our prediction on $\BR(\tau\to\mu\gamma)$.

The dark matter in this scenario is again the lightest $Z_2$-odd Majorana fermion $\psi_1$. The density of the thermal relic abundance depends on only $(Y_2^\dagger Y_2)_{11},\ M_{\psi_1}$ and $m_{\omega_2}$ in the case, where $Y_\eta$ and $\chi$ are negligibly small.
Values of these parameters are the same with those of Eq.(\ref{eq:benchmark_normal}).
Therefore, $M_{\psi_1}=97.6\ \GeV$ can still explain the observed relic density $\Omega_\mathrm{DM}h^2=0.1200\pm0.0012$~\cite{Aghanim:2018eyx},
just like in the benchmark scenario for the case of $m_1<m_3$.
The constraint from the direct detection experiments is also the same with the previous scenario.

\section{Conclusions}
\label{sec:Conclusions}
We have proposed a new mechanism to explain neutrino masses
with lepton number conservation,
in which the Dirac neutrino masses are
generated at the two-loop level
involving a dark matter candidate. In this model
branching ratios of lepton flavor violating decays of the Higgs boson
can be much larger than those of lepton flavor violating decays of
charged leptons. We have found the benchmark scenarios for normal ordered masses of neutrinos and inverted ones, where the neutrino mass matrix, the relic density of dark matter and the branching ratios for LFV processes can satisfy the constraints from current experimental data.
We have showed that $\BR(h\to\mu\tau)$ is about $10^2$ lager than $\BR(\tau\to\mu\gamma)$ in our benchmark scenarios. If the lepton flavor violating decays of the Higgs boson are observed
at the future collider experiments
without detecting lepton flavor violating decays of charged leptons,
most of the previously proposed models are excluded,
while our model can still survive.

In this paper, we did not discuss collider signature of new scalars and fermions.
Collider phenomenology for $Z_2$-even/odd charged singlet scalars in different models can be found in the literature~\cite{Kanemura:2000bq,Aoki:2010tf}/~\cite{Aoki:2010tf,Ahriche:2014xra, Aoki:2010aq}, while that for $\Phi$ $(Y=3/2)$ has been discussed in Ref.~\cite{Rentala:2011mr,Aoki:2011yk} in the different context.
We will discuss these issues elsewhere in the future~\cite{future_work}.

%%%%%%%%%%%%%%%%%%%%%%%%%%
%%%  Acknowledgments  %%%%
%%%%%%%%%%%%%%%%%%%%%%%%%%
\begin{acknowledgments}
 The work of K.~E.\ was supported in part by the Sasakawa Scientific Research Grant 
from The Japan Science Society.
 The work of S.~K.\ was supported in part
by Grant-in-Aid for Scientific Research on Innovative Areas,
the Ministry of Education, Culture, Sports, Science and Technology~(MEXT),
No.~16H06492,
No.~18H04587,
and Grant H2020-MSCA-RISE-2014 no.~645722~(Non Minimal Higgs).
 The work of K.~S.\ was supported in part
by Japan Society for the Promotion of Science~(JSPS) KAKENHI Grant
No.~18J12866~(JSPS Research Fellow).
 The work of H.~S.\ was supported in part
by MEXT KAKENHI Grant No.~18H05543~(Innovative Areas)
and JSPS KAKENHI Grant No.~18K03625~(Scientific Research~(C)).
\end{acknowledgments}

\newpage

%%%%%%%%%%%%%%%%%%
%%%  Appendix  %%%
%%%%%%%%%%%%%%%%%%
\appendix
\section{The loop function in the neutrino mass matrix}
\label{app:numass}
The neutrino mass matrix formula given by Eq.~(\ref{eq:neutrino_mass}) in Sec.~\ref{sec:Neutrino Mass} contains the loop function $I_{\ell' ak}$~$(k=1,2,\ a= 1,2,3~\text{and}~\ell^\prime = e,~\mu,~\tau)$. We here show the explicit formula for $I_{\ell' ak}$;
\bal
I_{\ell' a k}
=&
 \frac{ 1 }{ ( 16\pi^2 )^2 }\,
 \frac{1}
      {
       \bigl( m_{\pi_2}^2 - m_{\pi_1}^2 \bigr)
       \bigl( M^2_{\psi_a} - m_{\eta^0}^2 \bigr)
      }
\nonumber\\
&
 \times
 \int_0^1 \mathrm{d}z \hspace{2pt}
 z \biggl\{
     \frac{1}{ m_{\pi_1}^2 - m_{\ell'}^2 }
     \left(
      f_{ak}( m_{\pi_1}^2 )
      - f_{ak}( m_{\ell'}^2 )
     \right)
    -
     \frac{ 1 }{ m_{\pi_2}^2 - m_{\ell'}^2 }
     \left(
      f_{ak}( m_{\pi_2}^2 )
      - f_{ak}( m_{\ell'}^2 )
     \right)
   \biggr\},
\eal
where the function $f_{ak}$ is defined as follows
\bal
f_{ak}\left( m^2 \right)
=
 m^4
 \Bigl\{
   \mathrm{Li}_2 \bigl( z^\psi_{ka}\left( m^2 \right) \bigr)
  -
   \mathrm{Li}_2 \bigl( z^\eta_k\left( m^2 \right) \bigr)
 \Bigr\} ,
\eal
%--------
with
\bal
z^\psi_{ak}\left( m^2\right )
=&
  1
 -
  \frac{1}{ z \left( 1 - z \right) m^2 }
  \Bigl\{
    M_{\psi_a}^2
   +
    z \left( m^2_{\omega_k} - M^2_{\psi_a} \right)
  \Bigr\} ,
\\
%--------
z^\eta_k\left( m^2 \right)
=&
  1
 -
 \frac{1}{ z \left( 1 - z \right) m^2 }
 \Bigl\{
   m_{\eta^0}^2
  +
   z\left( m^2_{\omega_k} - m^2_{\eta^0} \right)
 \Bigr\} ,
\\
%---------
\mathrm{Li_2}(x)
=&
 \int_0^x \mathrm{d}t \hspace{2pt}
 \frac{1}{-t} \ln(1-t) .
\eal

\section{Some formulae for $\ell \to \ell^\prime \gamma$}
\label{app:ltolgamma}
In Sec.~\ref{sec:Lepton Flavor Violation}, blanching ratios for $\ell \to \ell' \gamma$ are given by Eq.~(\ref{eq:LFV-lepton-decay}), which depend on $A_R^{s_1},\ A_R^\omega$ and $A_L^\omega$. We here present their explicit formulae. They are given by
\bal
\label{eq:A1_cLFV}
A_R^{s_1}
&=
 \sum_k
 \frac{1}{12}
 \frac{ m^2_\ell }{ m_{\pi_k}^2 }
 \left( Y_1 Y_1^\dagger \right)_{\ell \ell'}
 \left( U_\theta \right)_{k2}^2
,
\\
%-------
\label{eq:A2_cLFV}
A_R^\omega
&=
 \sum_{a, k}
 \frac{1}{2}
 \frac{ m^2_\ell }{ m^2_{\omega_k} }
 \Biggl[
   \left( Y_2 \right)_{\ell' a}^\ast
   \left( Y_2 \right)_{\ell a}
   \left( U_\chi \right)_{k2}^2
   F_2\left( \frac{M_{\psi_a}^2}{m^2_{\omega_k}} \right)
\nonumber\\
&\hspace*{40mm}
{}-
   \frac{ M_{\psi_a} }{ m_\ell }
   \left( Y_2 \right)_{\ell' a}^\ast
   \left( Y_\eta \right)^\ast_{\ell a}
   \left( U_\chi \right)_{k1} \left( U_\chi \right)_{k2}
   G\left( \frac{M^2_{\psi_a}}{m^2_{\omega_k} } \right)
 \Biggr]
,
\\
%----------
\label{eq:B_cLFV}
A_L^\omega
&=
 \sum_{a, k}
 \frac{1}{2}
 \frac{ m^2_\ell }{ m^2_{\omega_k} }
 \Biggl[
   \left( Y_\eta \right)_{\ell' a}
   \left( Y_\eta \right)_{\ell a}^\ast
   \left( \chi'_k \right)^2
   F_2\left( \frac{M_{\psi_a}^2}{m^2_{\omega_k}} \right)
\nonumber\\
&\hspace*{40mm}
{}-
   \frac{ M_{\psi_a} }{ m_\ell}
   \left( Y_2 \right)_{\ell a}
   \left( Y_\eta \right)_{\ell' a}
   \left( U_\chi \right)_{k1} \left( U_\chi \right)_{k2}
   G\left( \frac{M^2_{\psi_a}}{m^2_{\omega_k}} \right)
 \Biggr]
,
\eal
where $F_2(x)$ and $G(x)$ are defined as 
\bal
%-----------
F_2(x)
&=\hspace{5pt}
 \frac{1}{ 6 ( 1 - x )^4 }
 \left(
  1 - 6x + 3x^2 + 2x^3 - 6x^2 \ln x
 \right) ,
\\
%-----------
G(x)
&=\hspace{5pt}
 \frac{1}{ ( 1 - x )^3 }
 \left(
  1 - x^2 + 2x \ln x
 \right)
.
\eal
 Terms that proportional to $M_{\psi_a}/m_\ell$
in formulae of $A_R^\omega$ and $A_L^\omega$
appear due to the mixing between $s_2^+$ and $\eta^+$.

\section{Some formulae for $ h \to \ell \ell^\prime$}
\label{app:htoll}
In Sec.~\ref{sec:Lepton Flavor Violation}, blanching ratios for $h \to \ell \ell'$ are given by Eq.~(\ref{eq:HiggsLFV}), which depend on $B_R^{s_1},\ B_R^\omega$ and $B_L^\omega$. We here give their explicit formulae. They are defined as 
\bal
B_R^{s_1}
=&
 m_\ell^{}
 \bigl( Y_1 Y_1^\dagger \bigr)_{\ell \ell^\prime}
 \sum_{k, k'}
 \Lambda_{k k'}^\pi
 \left( U_\theta \right)_{k 2}
 \left( U_\theta \right)_{k' 2}
 \int_0^1 \mathrm{d}x \mathrm{d}y \mathrm{d}z\
 \dfrac{ z }
       {
         y m_{\pi_k^{}}^2
        + 
         z m_{\pi_{k'}^{}}^2
        -
         y z m_h^2
       }
,
\\
%---------
B_R^\omega
=&
  \sum_{a, k, k'}
  m_\ell^{}
  \bigl( Y_2 \bigr)_{\ell a}
  \bigl( Y_2 \bigr)_{\ell^\prime a}^\ast
  \Lambda_{k k'}^\omega
  \left( U_\chi \right)_{k 2}
  \left( U_\chi \right)_{k' 2}
  \int_0^1 \mathrm{d}x \mathrm{d}y \mathrm{d}z\
  \dfrac{ z }
        {
          x M^2_{\psi_a}
         +
          y m^2_{\omega_k}
         +
          z m^2_{\omega_{k'}}
         -
          y z m^2_h
        }
\nonumber \\
&
{}+
  \sum_{a, k, k'}
  M_{\psi_a}
  \bigl( Y_\eta \bigr)_{\ell a}^\ast
  \bigl( Y_2 \bigr)_{\ell^\prime a}^\ast
  \Lambda_{k k'}^\omega
  \left( U_\chi \right)_{k 1}
  \left( U_\chi \right)_{k' 2}
  \int_0^1 \mathrm{d}x \mathrm{d}y \mathrm{d}z\
  \dfrac{ 1 }
        {
          x M^2_{\psi_a}
         +
          y m^2_{\omega_k}
         +
          z m^2_{\omega_{k'}}
         -
          y z m^2_h
        }
,
\\
%---------
B_L^\omega
=&
  \sum_{a, k, k'}
  m_\ell^{}
  \bigl( Y_\eta \bigr)_{\ell a}^\ast
  \bigl( Y_\eta \bigr)_{\ell^\prime a}
  \Lambda_{k k'}^\omega
  \left( U_\chi \right)_{k 1}
  \left( U_\chi \right)_{k' 1}
  \int_0^1 \mathrm{d}x \mathrm{d}y \mathrm{d}z\
  \dfrac{ z }
        {
          x M^2_{\psi_a}
         +
          y m^2_{\omega_k}
         +
          z m^2_{\omega_{k'}}
         -
          y z m^2_h
        }
\nonumber \\
&
{}+
  \sum_{a, k, k'}
  M_{\psi_a}
  \bigl( Y_2 \bigr)_{\ell a}
  \bigl( Y_\eta \bigr)_{\ell^\prime a}
  \Lambda_{k k'}^\omega
  \left( U_\chi \right)_{k 2}
  \left( U_\chi \right)_{k' 1}
  \int_0^1 \mathrm{d}x \mathrm{d}y \mathrm{d}z\
  \dfrac{ 1 }
        {
          x M^2_{\psi_a}
         +
          y m^2_{\omega_k}
         +
          z m^2_{\omega_{k'}}
         -
          y z m^2_h
        }
.
\eal

Coefficients $\Lambda^\pi_{kk'}$ and $\Lambda^\omega_{kk'}$ are defined in order to satisfy
\bal
\mathcal{L}
=&
 \sum_{k, k'}
 \left(
   \Lambda_{k k'}^\pi
   \pi_k^+ \pi_{k'}^-
  +
   \Lambda_{k k'}^\omega
   \omega_k^+ \omega_{k'}^-
 \right)
 h
,
\eal
and given by
\bal
%----------
\Lambda_{11}^\pi
=&
 -
  \dfrac{ \sigma_1 }{ \sqrt{2} }
  \sin{2\theta}
 -
  \bigl(
    \lambda_{\phi\Phi}
   +
    \lambda'_{\phi\Phi}
  \bigr)
  v \cos^2{\theta}
 -
  \lambda_{\phi1}
  v \sin^2{\theta}
,
\\
%----------
\Lambda_{12}^\pi
=
 \Lambda_{21}^\pi
=&
 -
  \dfrac{ \sigma_1 }{ \sqrt{2} }
  \cos{2\theta}
 +
  \frac{1}{2}
  \bigl(
    \lambda_{\phi\Phi}
   +
    \lambda'_{\phi\Phi}
  \bigr)
  v \sin{2\theta}
 -
  \frac{1}{2}
  \lambda_{\phi1}
  v \sin{2\theta}
,
\\
%--------
\Lambda_{22}^\pi
=&
  \dfrac{ \sigma_1 }{ \sqrt{2} }
  \sin{2\theta}
 -
  \bigl(
    \lambda_{\phi\Phi}
   +
    \lambda'_{\phi\Phi}
  \bigr)
  v \sin^2{\theta}
 -
  \lambda_{\phi1}
  v \cos^2{\theta}
,
\\
%---------
\Lambda_{11}^\omega
=&
  \dfrac{ \sigma_3 }{ \sqrt{2} }
  \sin{2\chi}
 -
  \lambda_{\phi\eta}
  v \cos^2{\chi}
 -
  \lambda_{\phi2}
  v \sin^2{\chi}
,
\\
%---------
\Lambda_{12}^\omega
=
 \Lambda_{21}^\omega
=&
  \dfrac{ \sigma_3 }{ \sqrt{2} }
  \cos{2\chi}
 +
  \frac{1}{2}
  \lambda_{\phi\eta}
  v \sin{2\chi}
 -
  \frac{1}{2}
  \lambda_{\phi2}
  v \sin{2\chi}
,
\\
%--------
\Lambda_{22}^\omega
=&
 -
  \dfrac{ \sigma_3 }{ \sqrt{2} }
  \sin{2\chi}
 -
  \lambda_{\phi\eta}
  v \sin^2{\chi}
 -
  \lambda_{\phi2}
  v \cos^2{\chi}
.
\eal

\section{Some formulae for $\ell_m\to\overline{\ell}_n\ell_p\ell_q$}
\label{app:ltolll}
In Sec.~\ref{sec:Lepton Flavor Violation}, blanching ratios for $\ell_m\to\overline{\ell}_n\ell_p\ell_q$ are given by Eq.~(\ref{eq:LFV_decay_via_box_diagram}), which depend on $(C_{RRRR}^{s_1})_{mnpq},\ (C_{RRRR}^{s_2})_{mnpq},\ (C_{}^{\eta})_{mnpq}$ and $(C^\omega)_{mnpq}{}'\mathrm{s}$. We here give their explicit formulae. They are given by
\bal
\left( C_{RRRR}^{s_1} \right)_{mnpq}
=&
 -
 \frac{1}{2}
 \Bigl[
   \left( Y_1 Y_1^\dagger \right)_{mp}
   \left( Y_1 Y_1^\dagger \right)_{nq}
  +
  (m\leftrightarrow n)
 \Bigr]
 \sum_{k, k'}
 \left( U_\theta \right)_{k 2}^2
 \left( U_\theta \right)_{k' 2}^2
 \int
 \frac{ 1 }
      { \Delta }
,
\\
%----------
\left( C_{RRRR}^{s_2} \right)_{mnpq}
=&
 -
  \sum_{a, b, k, k'}
  \frac{1}{2}
  \Bigl(
    \left( Y_2 \right)_{m a}
    \left( Y_2 \right)_{n b}
    \left( Y_2 \right)_{p a}^\ast
    \left( Y_2 \right)_{q b}^\ast
   +
    \left( p \leftrightarrow q \right)
  \Bigr)
  \left( U_\chi \right)_{k 2}^2
  \left( U_\chi \right)_{k' 2}^2
  \int
  \frac{ 1 }
       { \Sigma }
\nonumber \\
&
{}-
  \sum_{a, b, k, k'}
  M_{\psi_a} M_{\psi_b}
  \left( Y_2 \right)_{m a}
  \left( Y_2 \right)_{n a}
  \left( Y_2 \right)_{p b}^\ast
  \left( Y_2 \right)_{q b}^\ast
  \left( U_\chi \right)_{k 2}^2
  \left( U_\chi \right)_{k' 2}^2
  \int
  \frac{ 1 }
       { \Sigma ^2 }
,
\\
%------------
\left( C_{LLLL}^\eta \right)_{mnpq}
=&
 -
  \sum_{a, b, k, k'}
  \frac{1}{2}
  \Bigl(
    \left( Y_\eta \right)_{m a}^\ast
    \left( Y_\eta \right)_{n b}^\ast
    \left( Y_\eta \right)_{p a}
    \left( Y_\eta \right)_{q b}
   +
    \left( p \leftrightarrow q \right)
  \Bigr)
  \left( U_\chi \right)_{k 1}^2
  \left( U_\chi \right)_{k' 1}^2
  \int \frac{ 1 }{ \Sigma }
\nonumber \\
&
{}-
  \sum_{a, b, k, k'}
  M_{\psi_a}
  M_{\psi_b}
  \left( Y_\eta \right)_{m a}^\ast
  \left( Y_\eta \right)_{n a}^\ast
  \left( Y_\eta \right)_{p b}
  \left( Y_\eta \right)_{q b}
  \left( U_\chi \right)_{k 1}^2
  \left( U_\chi \right)_{k' 1}^2
  \int \frac{ 1 }{ \Sigma^2 }
,
\\
%--------
\left( C_{LLRR}^\omega \right)_{mnpq}
=&
 \sum_{a, b, k, k'}
 M_{\psi_a}
 M_{\psi_b}
 \left( Y_\eta \right)_{m a}^\ast
 \left( Y_\eta \right)_{n b}^\ast
 \left( Y_2    \right)_{p a}^\ast
 \left( Y_2    \right)_{q b}^\ast
 \left( U_\chi \right)_{k 1}
 \left( U_\chi \right)_{k' 1}
 \left( U_\chi \right)_{k 2}
 \left( U_\chi \right)_{k' 2}
 \int \frac{1}{\Sigma^2}
,
\\
\left( C_{RRLL}^\omega \right)_{mnpq}
=&
 \sum_{a, b, k, k'}
 M_{\psi_a}
 M_{\psi_b}
 \left( Y_2    \right)_{m a}
 \left( Y_2    \right)_{n b}
 \left( Y_\eta \right)_{p a}
 \left( Y_\eta \right)_{q b}
 \left( U_\chi \right)_{k 2}
 \left( U_\chi \right)_{k' 2}
 \left( U_\chi \right)_{k 1}
 \left( U_\chi \right)_{k' 1}
 \int \frac{ 1 }{ \Sigma^2 }
,
\eal
\bal
\left( C_{RLLR}^\omega \right)_{mnpq}
=&
 \sum_{a, b, k, k'}
 \left( Y_2    \right)_{m a}
 \left( Y_\eta \right)_{n b}^\ast
 \left( U_\chi \right)_{k 2}
 \left( U_\chi \right)_{k' 1}
\nonumber\\
&\hspace*{10mm}
 \times
 \Bigl(
   M_{\psi_a} M_{\psi_b}
   \left( Y_\eta \right)_{p a}
   \left( Y_2    \right)_{q b}^\ast
   \left( U_\chi \right)_{k' 1}
   \left( U_\chi \right)_{k 2}
   \int \frac{ 1 }{ \Sigma^2 }
  +
   \left( Y_\eta \right)_{p b}
   \left( Y_2    \right)_{q a}^\ast
   \left( U_\chi \right)_{k 1}
   \left( U_\chi \right)_{k' 2}
   \int \frac{ 1 }{ \Sigma }
 \Bigr)
\nonumber \\
&
{}-
  \sum_{a, b, k, k'}
  \left( Y_2    \right)_{m a}
  \left( Y_\eta \right)_{n a}^\ast
  \left( U_\chi \right)_{k 2}
  \left( U_\chi \right)_{k' 1}
  \left( Y_\eta \right)_{p b}
  \left( Y_2    \right)_{q b}^\ast
  \nonumber\\
&\hspace*{10mm}
 \times
  \biggl(
    \left( U_\chi \right)_{k 1}
    \left( U_\chi \right)_{k' 2}
   -
    \left( U_\chi \right)_{k' 1}
    \left( U_\chi \right)_{k 2}
  \biggr)
  \int \frac{ 1 }{ \Sigma }
,
\\
%----------
\left( C_{LRRL}^\omega \right)_{mnpq}
=&
 \sum_{a, b, k, k'}
 \left( Y_\eta \right)_{m a}^\ast
 \left( Y_2    \right)_{n b}
 \left( U_\chi \right)_{k 1}
 \left( U_\chi \right)_{k' 2}
\nonumber\\
&\hspace*{10mm}
 \times
 \Bigl(
   M_{\psi_a} M_{\psi_b}
   \left( Y_2    \right)_{p a}^\ast
   \left( Y_\eta \right)_{q b}
   \left( U_\chi \right)_{k 1}
   \left( U_\chi \right)_{k' 2}
   \int \frac{ 1 } { \Sigma^2 }
  +
   \left( Y_2    \right)_{p b}^\ast
   \left( Y_\eta \right)_{q a}
   \left( U_\chi \right)_{k 2}
   \left( U_\chi \right)_{k' 1}
   \int \frac{ 1 }{ \Sigma }
 \Bigr)
\nonumber \\
&
{}-
  \sum_{a, b, k, k'}
  \left( Y_\eta \right)_{m a}^\ast
  \left( Y_2    \right)_{n a}
  \left( U_\chi \right)_{k 1}
  \left( U_\chi \right)_{k' 2}
  \left( Y_2    \right)_{p b}^\ast
  \left( Y_\eta \right)_{q b}
\nonumber\\
&\hspace*{10mm}
 \times
  \Bigl(
    \left( U_\chi \right)_{k 2}
    \left( U_\chi \right)_{k' 1}
   -
    \left( U_\chi \right)_{k 1}
    \left( U_\chi \right)_{k' 2}
  \Bigr)
  \int \frac{ 1 }{ \Sigma }
,
\eal
where $\Delta$ and $\Sigma$ are defined as 
\bal
\Delta
&=
  x m^2_{\pi_k}
 +
  y m^2_{\pi_{k'}}
\\
%---------
\Sigma
&=
  x m^2_{\omega_k}
 +
  y m^2_{\omega_{k'}}
 +
  z M^2_{\psi_a}
 +
  \omega M^2_{\psi_b}\ ,
\eal
and the symbol  $\int$ denotes the integration with respect to $x, y, z$ and $\omega$ as follows; 
\bal
\int=\int_0^1\ \mathrm{d}x\hspace{1pt}\mathrm{d}y\hspace{1pt}\mathrm{d}z\hspace{1pt}\mathrm{d}\omega \ .
\eal

By exchanging $p$ and $q$ 
for $\left( C_{RLLR}^\omega \right)_{mnpq}$
and $\left( C_{LRRL}^\omega \right)_{mnpq}$,
we obtain
\bal
\left( C_{RLRL}^\omega \right)_{mnpq}
=&
 - \left( C_{RLLR}^\omega \right)_{mnqp}
,
\\
%----------
\left( C_{LRLR}^\omega \right)_{mnpq}
=&
 - \left( C_{LRRL}^\omega \right)_{mnqp}
.
\eal

\section{ Annihilation of dark matter $\psi_a$ }
\label{app:DM}
In Sec.~\ref{sec:Dark Matter}, we have shown only the approximate formula for the thermal averaged cross section for annihilation of the dark matter $\psi_a$, $\left<\sigma v_\mathrm{rel}\right>$. In this appendix, we show the complete formula at tree level. First, the contribution from annihilation to a pair of charged leptons, $\left<\sigma_\ell v_\mathrm{rel}\right>$, which is shown by the left of Fig.~\ref{fig:relic_abundance}, is given by 
\bal
\left< \sigma_\ell^{} v_\mathrm{rel} \right>
=&
  \sum_{k, k'}
  \frac{ 1 }{ 8\pi }
  \Bigl(
    \bigl( Y_2^\dagger Y_2 \bigr)_{aa}^2
    \left( U_\chi \right)_{k 2}^2
    \left( U_\chi \right)_{k' 2}^2
   +
    \bigl( Y_\eta^\dagger Y_\eta \bigr)_{aa}^2
    \left( U_\chi \right)_{k 1}^2
    \left( U_\chi \right)_{k' 1}^2
  \Bigr)
  \nonumber \\
&\hspace{15pt}
  \times
  \frac{
        M_{\psi_a}^2
        \bigl(
          M^4_{\psi_a}
         +
          m^2_{\omega_k}
          m^2_{\omega_{k'}}
        \bigr)
       }
       {
        ( M^2_{\psi_a} + m^2_{\omega_k} )^2
         ( M^2_{\psi_a} + m^2_{\omega_{k'}} )^2 
       }\,
  \frac{ 1 }{\,x\,}
\nonumber \\
&
{}+
  \sum_{k, k'}
  \frac{ 1 }{ 16\pi }
  \bigl( Y_2^\dagger Y_2 \bigr)_{aa}
  \bigl( Y_\eta^\dagger Y_\eta \bigr)_{aa}
  \left( U_\chi \right)_{k 2}^2
  \left( U_\chi \right)_{k' 2}^2
  \left( U_\chi \right)_{k 1}^2
  \left( U_\chi \right)_{k' 1}^2
\nonumber\\
&\hspace*{15pt}
  \times
  \Biggl[
    \frac{ 2 M_{\psi_a}^2 }
         {
          ( M^2_{\psi_a} + m^2_{\omega_k} )
          ( M^2_{\psi_a} + m^2_{\omega_{k'}} )
         }
  +
    \frac{ M^2_{\psi_a} }
         {
          ( M^2_{\psi_a} + m^2_{\omega_k} )^3
          ( M^2_{\psi_a} + m^2_{\omega_{k'}} )^3
         }
\nonumber \\
&\hspace{12mm}
    \times
    \Bigl\{
      5 M_{\psi_a}^8
     +
      12 M^6_{\psi_a}
         ( m^2_{\omega_k} + m^2_{\omega_{k'}} )
     +
      3 M^4_{\psi_a}
        (
          m^4_{\omega_k}
         +
          8 m^2_{\omega_k} m^2_{\omega_{k'}}
         +
          m^4_{\omega_{k'} }
        )
\nonumber \\
&\hspace{45mm}
{}   +
      4 M^2_{\psi_a} m^2_{\omega_k} m^2_{\omega_{k'}}
        ( m^2_{\omega_k} + m^2_{\omega_{k'}} )
     -
      3 m^4_{\omega_k} m^4_{\omega_{k'}}
    \Bigr\}\,
    \frac{ 1 }{\,x\,}
  \Biggr] .
 \label{eq:Full_Formula_for_sigmav_1}
\eal

Second, the contribution from annihilation to a pair of neutrinos, $\left<\sigma_\nu v_\mathrm{rel}\right>$, which is represented by the right of Fig.~\ref{fig:relic_abundance}, is given by
\bal 
\left<\sigma_\nu v_\mathrm{rel}\right> = 
\frac{
       \left(Y_\eta^\dagger Y_\eta\right)_{aa}^2 }
      { 
       8\pi
       }
  \frac{ 
          M_{\psi_a}^2\left(M^4_{\psi_a}+m_{\eta}^4\right)
          }
         {
          \left(M_{\psi_a}^2+m_\eta^2\right)^4
         }
  \frac{ 1 }{ x }.
  \label{eq:Full_Formula_for_sigmav_2}
\eal
The complete formula for $\left<\sigma v_\mathrm{rel}\right>$ is given  at tree level by the sum of Eq.~(\ref{eq:Full_Formula_for_sigmav_1}) and~(\ref{eq:Full_Formula_for_sigmav_2}).

\end{document}